\documentclass[preprint,showpacs,preprintnumbers,amsmath,amssymb]{revtex4}

\usepackage{graphicx}
\usepackage{dcolumn}
\usepackage{bm}

\begin{document}


\title{Intrinsic thermodynamic properties of the pyrochlore superconductor RbOs$_2$O$_6$ extracted by condensation energy analysis}

\author{M.~Br\"uhwiler}
 \email{bruehwiler@solid.phys.ethz.ch}
 \affiliation{Laboratory for Solid State Physics, ETH Z\"urich, 8093 Z\"urich, Switzerland.}
\author{S.M.~Kazakov}%
 \affiliation{Laboratory for Solid State Physics, ETH Z\"urich, 8093 Z\"urich, Switzerland.}
\author{J.~Karpinski}%
 \affiliation{Laboratory for Solid State Physics, ETH Z\"urich, 8093 Z\"urich, Switzerland.}
\author{B.~Batlogg}%
 \affiliation{Laboratory for Solid State Physics, ETH Z\"urich, 8093 Z\"urich, Switzerland.}

\date{\today}

\newcommand{\fu}{$\mathrm{RbOs_2O_6}$}
\newcommand*{\unit}[1]{\,\mathrm{#1}}
\newcommand{\ce}{\Delta U}  
\newcommand{\CE}{\widetilde{\ce}}  
\newcommand{\sfc}{\gamma}  
\newcommand{\SFC}{\widetilde{\sfc}}  
\newcommand{\Tc}{T_\mathrm{c}}  
\newcommand{\Cp}{C_\mathrm{p}}  
\newcommand{\Hc}{H_\mathrm{c}}  
\newcommand{\svf}{superconducting volume fraction}
\newcommand{\smf}{superconducting mass fraction}

\begin{abstract}

We develop a general procedure for the analysis of bulk
thermodynamic data of a superconductor for samples containing a
metallic non-superconducting second phase. The method is based on
the condensation energy and it allows the extraction of the
intrinsic properties of a superconductor even for non-ideal
samples. Applying this procedure to the recently discovered
geometrically frustrated $\beta$-pyrochlore superconductor \fu\
($\Tc=6.4\unit{K}$) yields a Sommerfeld coefficient as high as
$79\unit{\mu J/g/K^2}$ ($44\unit{mJ/mol_\mathrm{f.u.}/K^2}$). \fu\
is inferred to be a strong type-II superconductor
($\kappa(\Tc)=23$) in the intermediate-coupling regime similar to
niobium ($\lambda_\mathrm{ep} \approx 1$). From the upper critical
field $\mu_0 H_{\mathrm{c2}} \approx 6\unit{T}$ at $0\unit{K}$, we
estimate a Ginzburg-Landau coherence length $\xi \approx
74\unit{\AA}$. The condensation energy is $860\unit{\mu J/g}$
($483\unit{mJ/mol_\mathrm{f.u.}}$) resulting in $1/(8\pi) \cdot
(\sfc_1\Tc^2)/\ce_1 \approx 0.15$, a value well in the range of
conventional phonon-mediated superconductors. The superconducting
electronic specific heat indicates conventional s-wave pairing.
The experimental Sommerfeld coefficient of
$44\unit{mJ/mol_\mathrm{f.u.}/K^2}$ is about $4$ times larger than
the one found in band structure calculations. Together with the
electron-phonon coupling constant $\lambda_\mathrm{ep} \approx 1$
this leaves an additional $\lambda_\mathrm{add} \approx 2.4$ for
enhancement due to other mechanisms.
\end{abstract}

\pacs{74.25.Op, 74.25.Bt, 74.70.-b}

\maketitle


\section{Introduction}

Thermodynamic studies of single phase superconductor samples are
most desirable, yet it often proves difficult to synthesize
samples which are $100\unit{\%}$ pure. Especially in the initial
phase of the discovery of a new superconducting material,
synthesis methods usually are not as advanced as to synthesize
perfect samples. At this stage a significant amount of a precursor
material or a by-product often remains as a second component in
the end product. Therefore, much care must be taken when
extracting properties of the superconductor to ensure that these
are intrinsic.

For meaningful physical statements, intensive quantities need to
be evaluated, i.e.~the measured extensive physical quantities need
to be given per amount of substance (a.o.s.), per volume, or per
mass. However, since the ratio of the two components is not known,
or the second component is considered to contribute marginally,
the quantities are usually given per \textit{total} a.o.s.,
volume, or mass. Values calculated in this way can be
significantly off the true value and require a more careful
analysis.

The key issue therefore is extracting the fraction of each
component present. When dealing with a superconducting sample
containing a second \textit{metallic} component, such an analysis
is possible and is described here. Integrating the heat capacity
to obtain the condensation energy of the superconductor gives a
measure for the amount of the superconducting component. When
several samples of different superconducting volume fractions are
measured, the condensation energy obtained varies accordingly.
From the systematic variation it is possible to specify the
properties of the ideal sample.

\begin{figure}
  \includegraphics{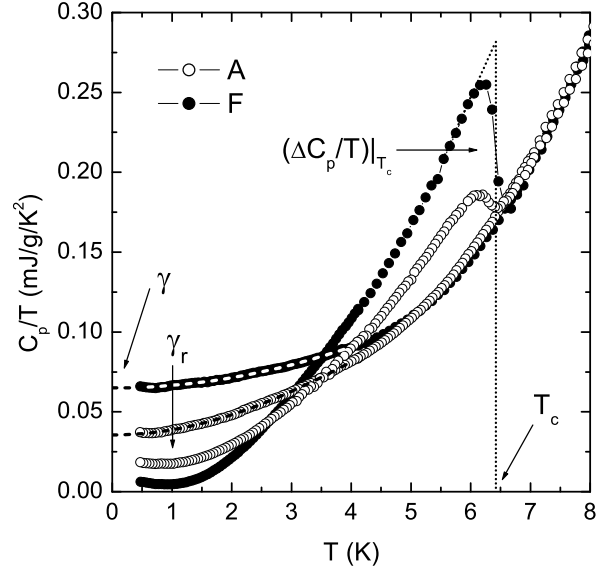}
  \caption{\label{fig:CdivT_vs_T} Specific heat divided by
  temperature $\Cp/T$ for samples A and F at $0\unit{T}$ and
  $12\unit{T}$. We extract the Sommerfeld coefficient $\sfc$ by fitting the data from
$1\unit{K}$ to $4\unit{K}$ to $C_{\mathrm{p}}/T = \gamma + \beta
T^2 + D T^4$ (shown by dashed lines). The residual coefficient
$\sfc_\mathrm{r}$ is the $\Cp/T$ value at $1\unit{K}$ and
$0\unit{T}$, and $\Tc$ is chosen such as to balance the entropy.
The residual Sommerfeld coefficient $\sfc_\mathrm{r}$, the
normal-state coefficient $\sfc$, and the specific heat jump at
$\Tc$, $\Delta \Cp \vert_{\Tc}/\Tc$, vary among the various
samples measured in a way that is a typical signature of a
metallic second component. These observations form the basis for
the condensation energy analysis (CEA) presented here.}
\end{figure}

Specific heat measurements on \fu\ have shown that in the
superconducting state a noticeable density of states at the Fermi
level remains as $T \to 0\unit{K}$ (Fig.~\ref{fig:CdivT_vs_T}).
The residual Sommerfeld coefficient $\sfc_\mathrm{r}$ and
normal-state coefficient $\sfc$ vary among the various synthesis
runs in a way that is a typical signature of the above described
situation. We have therefore developed the appropriate analysis
method described below.

\fu\ is one of only four pyrochlore superconductors known to date,
of which three were discovered very recently and belong to the
same family. These are the $\alpha$-pyrochlore Cd$_2$Re$_2$O$_7$
\cite{HaMuTaSaYaHi2001,SaYoOhKaKaWaMaHaOn2001,JiHeMcAlDrMa2001}
and the $\beta$-pyrochlores \textit{A}Os$_2$O$_6$, where
\textit{A} = Cs, Rb, or K
\cite{YoMuHi2004,YoMuMaHi2004a,YoMuMaHi2004}. The pyrochlores
exhibit inherent geometrical frustration due to the metal ions
forming a network of corner-sharing tetrahedra. Thus, metallic
pyrochlores constitute ideal systems to study to what degree
itinerant electrons are affected by a lattice which is known to
cause geometrical frustration for interactions of localized
magnetic moments.

As reported previously \cite{BrKaZhKaBa2004}, \fu\ is a
conventional s-wave superconductor with a critical temperature
$\Tc = 6.4\unit{K}$. This analysis was based on specific heat
measurements and evidence for s-wave symmetry was also given by
penetration depth measurements by
\citeauthor{KhEsKaKaZhBrGaDiShMaMaBaKe2004}
(Ref.~\cite{KhEsKaKaZhBrGaDiShMaMaBaKe2004}). Further support for
conventional s-wave-type superconductivity in \fu\ comes from
${}^{87}$Rb NMR measurements by \citeauthor{MaGaPeHiWeOtKaKa2004}
\cite{MaGaPeHiWeOtKaKa2004}. Here, we shed new light onto this
subject by applying the developed analysis to our data on \fu. In
particular, we provide evidence for an additional electronic mass
enhancement beyond the contribution from electron-phonon coupling.

This article is divided into two parts: First, the condensation
energy analysis is described and appropriate formulas derived. In
the second part we apply the analysis to \fu\ to extract its
intrinsic thermodynamic parameters. For illustration purposes, we
frequently refer to the Figures pertaining to \fu\ already in the
general section. Finally, we elaborate on some further aspects of
the condensation energy analysis in the Appendices.

\section{Condensation energy analysis}

The starting point for our analysis is the experimentally measured
extensive heat capacity $\widetilde{C}_\mathrm{p}$ of a sample
\footnote{We use a tilde ($\widetilde{\hspace{1em}}$) to denote
quantities in an extensive form. Often, a capital $C$ is used to
denote (extensive) heat capacity, while a small $c$ is used to
denote (intensive) specific heat, i.e.~$c=C/m$. For several
reasons, this notation is not appropriate here. First, we also use
heat capacity per volume and a.o.s., for which there is no
dedicated symbol. Second, the same difficulty of an
extensive/intensive quantity applies to the Sommerfeld coefficient
$\sfc$, for which there is only one customary symbol. We therefore
feel that a tilde is a more advisable way to distinguish extensive
from intensive quantities. Further, the terms "specific heat" and
"heat capacity" are often used interchangeably and we therefore
prefer to use the terms intensive/extensive heat capacity
instead.}. We assume that two components contribute to this heat
capacity: component \textit{one} is the superconductor, while
component \textit{two} is a normal metal:
\begin{equation}\label{eqn:hc_definition}
  \widetilde{C}_\mathrm{p} = \widetilde{C}_1 + \widetilde{C}_2.
\end{equation}
The fact that the first component becomes superconducting enables
us to extract the contribution to the heat capacity from each
component by analyzing the condensation energy. We therefore call
this procedure condensation energy analysis (CEA).

Because the relative contribution of each component is not known a
priori, and because a sample dependent extensive heat capacity is
not meaningful, an experimenter assumes the contribution
$\widetilde{C}_2$ of component two to be small and uses the total
mass $m$ of the sample to convert the heat capacity into an
intensive form $\Cp = \widetilde{C}_\mathrm{p}/m$. In this case,
the heat capacity $\Cp$ is given in energy per temperature per
mass (typically in $\mu$J/g/K). Due to the presence of a second
component, this heat capacity is obviously not the correct heat
capacity of component one.

Usually the total mass of a sample is measured and conversion into
an intensive form is performed using this mass. Consequently, in
our analysis we will also assume the total mass to be known
experimentally. As will be shown below, this results in the
properties of the sample to be linear in the superconducting mass
fraction.

The CEA not only allows the extraction of the intrinsic properties
of the superconducting phase, it also does so without requiring
any further knowledge about the second normal metallic component.
We use the presence of varying amounts of this second component to
specify the properties of the ideal sample: The heat capacity
$\Cp/T$ for the superconductor vanishes for $T \ll \Tc$, while the
measured residual heat capacity is due to the second, metallic
phase only. To a certain extent, it is possible to extract
intrinsic thermodynamic properties of the second component as
well. If we can actually identify component two, e.g.~from XRD
structure analysis, this extraction is possible to the full
extent, as we show in the present example.

\subsection{Definitions}

The following definitions will prove useful in making formulas
more readable:
\begin{equation}\label{eqn:ratios_1}
  \mu := \frac{\rho_1}{\rho_2} \qquad
  \epsilon := \frac{\sfc_2}{\sfc_1}.
\end{equation}
Here, $\rho_1$ and $\rho_2$ are the mass densities of the two
components in mass per volume and $\mu$ is their ratio. $\sfc_1$
and $\sfc_2$ are the Sommerfeld coefficients of the two components
in energy per mass per temperature squared (typically in
$\mu$J/g/K$^2$) and $\epsilon$ is their ratio.

We further assign $\eta_V$ to the superconducting volume fraction
and $\eta_m$ to the superconducting mass fraction:
\begin{equation}\label{eqn:ratios_2}
  \eta_V := \frac{V_1}{V}, \qquad  \eta_m := \frac{m_1}{m}.
\end{equation}
$V_i$ and $m_i$ are the volume and mass of component $i$, while
$V$ and $m$ are the total volume and total mass respectively,
i.e.~$V_1 + V_2 = V$ and $m_1 + m_2 = m$.

\subsection{Parameters extracted from the measurement}

The parameters which are readily extracted from specific heat
measurements and also published as such are the Sommerfeld
coefficients of the two-component system $\sfc$ and
$\sfc_\mathrm{r}$ in the normal state and the superconducting
state respectively. We call the Sommerfeld coefficient in the
superconducting state $\sfc_\mathrm{r}$ "residual", because it
results from the second component only, since the first one
becomes superconducting with a vanishing heat capacity for $T \to
0$. We calculate these as a function of the superconducting mass
fraction $\eta_m$, which is proportional to the condensation
energy $\ce = \eta_m \cdot \ce_1$. $\ce_1 = \ce\vert_{\eta_m=1}$
is the condensation energy of the superconducting component. $\ce$
for a given sample is readily obtained by integrating the measured
specific heat to obtain the entropy difference
\begin{equation}\label{eqn:DeltaS_of_T}
  \Delta S (T) = \int_{T}^{\Tc} \frac{\Delta \Cp(T')}{T'} \mathrm{d}T'
\end{equation}
and then integrating again to obtain
\begin{equation}\label{eqn:Hc_of_T}
  \ce(T) = \int_{T}^{\Tc} \Delta S(T') \mathrm{d}T'.
\end{equation}
Here, $\Delta \Cp := C_{0\unit{T}} - C_{12\unit{T}}$ is the
difference in heat capacity between the superconducting and normal
state. In the present study, the normal state is reached by
applying an external magnetic field of $12\unit{T}$, which is well
above the upper critical field. The thermodynamic critical field
$\Hc$ is obtained from the condensation energy by using the
relation $\frac{1}{2} \mu_0 \Hc^2(T) = \ce_1(T) \rho_1 =
\eta_m^{-1} \ce(T) \rho_1$.

The relations between the intrinsic Sommerfeld coefficients of the
two components, $\sfc_1$ and $\sfc_2$, and the measured Sommerfeld
coefficients $\sfc$ and $\sfc_\mathrm{r}$ are most easily obtained
from their respective extensive forms:

The residual Sommerfeld coefficient $\SFC_\mathrm{r}$ measured in
$0\unit{T}$ is equivalent to the Sommerfeld coefficient $\SFC_2$
of the second component, since the density of states of the
superconducting component is gapped and therefore $\sfc_1 =0$:
\begin{equation}\label{eqn:gamma_res_a}
  m \sfc_\mathrm{r} = \SFC_\mathrm{r} = \SFC_2 = m_2 \sfc_2.
\end{equation}
From this we get
\begin{equation}\label{eqn:gamma_res_b}
  \sfc_\mathrm{r} = - \sfc_2 \eta_m + \sfc_2.
\end{equation}
The residual Sommerfeld coefficient $\sfc_\mathrm{r}$ measured in
$0\unit{T}$ is a direct measure for the presence of the second
component. It is easily seen that the $y$-intercept of a
$\sfc_\mathrm{r}$ vs $\eta_m$ plot is equal to $\sfc_2$ and that
$\sfc_\mathrm{r}$ vanishes at $\eta_m = 1$ and thus defines
$100\unit{\%}$ superconducting volume or mass fraction.

To calculate $\sfc$ we equate
\begin{equation}\label{eqn:gamma_a}
  m \sfc = \SFC = \SFC_1 + \SFC_2 = m_1 \sfc_1 + m_2 \sfc_2,
\end{equation}
resulting in
\begin{equation}\label{eqn:gamma_b}
  \sfc = (\sfc_1 - \sfc_2) \eta_m + \sfc_2,
\end{equation}
which also depends linearly on the \smf\ $\eta_m$. Thus, from
$\sfc$ vs $\eta_m$, we get a second, independent plot to extract
$\sfc_2$ by noting the $y$-intercept. Having located the ideal
sample through $\sfc_\mathrm{r}$, the intrinsic Sommerfeld
coefficient of component one, $\sfc_1$, can immediately be
extracted by taking the value of $\sfc$ at $\eta_m = 1$.

\subsection{Superconducting electronic specific heat}

The superconducting electronic specific heat $C_\mathrm{es}$ is
extracted in the usual way by subtracting the specific heat at
$0\unit{T}$ from the normal state specific heat at $12\unit{T}$
but paying attention to the appropriate masses:
\begin{eqnarray}\label{eqn:C_12T_0T}
  \widetilde{C}_{0\unit{T}} = m C_{0\unit{T}} = m_1 C_\mathrm{es} + \SFC_2 T + \widetilde{C}_\mathrm{lattice} +
  \widetilde{C}_\mathrm{other} \\
  \widetilde{C}_{12\unit{T}} = m C_{12\unit{T}} = \SFC_1 T + \SFC_2 T + \widetilde{C}_\mathrm{lattice} +
  \widetilde{C}_\mathrm{other} + \widetilde{C}_\mathrm{mag}.
\end{eqnarray}
Here, $C_\mathrm{lattice}$, $C_\mathrm{other}$, and
$C_\mathrm{mag}$ are the total heat capacity of both components
from the respective source. $C_\mathrm{mag}$ describes any
field-induced heat capacity, be it from fluctuations or from a
magnetization of the material. Taking the difference and assuming
any magnetic heat capacity $C_\mathrm{mag}$ to be negligible, we
get
\begin{equation}\label{eqn:Ces}
  C_\mathrm{es}(T) = \eta_m^{-1} \Delta \Cp +
  \sfc_1 T.
\end{equation}
We conclude that the specific heat difference $\Delta \Cp$ needs
to be scaled with the inverse \smf\ to extract the true electronic
superconducting specific heat. Also, the added electronic specific
heat from the normal state, $\sfc_1 T$, has to be calculated using
the Sommerfeld coefficient $\sfc_1$ of the pure superconductor.
For an ideal sample with $\eta_m = 1$, Eq.~(\ref{eqn:Ces}) is, of
course, equivalent to the usual $C_\mathrm{es}(T) = \Delta \Cp +
\sfc T$. As Figure \ref{fig:Ces_vs_TdivTc} shows for \fu, the
electronic superconducting specific heat indeed shows a universal
behavior when scaled appropriately according to
Eq.~(\ref{eqn:Ces}).

At low temperatures, where $C_\mathrm{es} \ll \sfc_1 T$, we can
approximate $\Delta \Cp$ using Eq.~(\ref{eqn:Ces}) and get
\begin{equation}\label{eqn:DeltaC}
  \Delta \Cp = - \sfc_1 \eta_m T.
\end{equation}
That is, $\Delta \Cp$ is linear in $T$ and the slope is
proportional to $\eta_m$, varying from sample to sample. This
behavior is illustrated in Fig.~\ref{fig:DeltaC_vs_T} for \fu.
Evidently, the negative of the mentioned slope plotted against the
\smf\ has the value $\sfc_1$ at $\eta_m=1$. $\Delta \Cp$ thus
provides another way of extracting the true Sommerfeld coefficient
of component one from the data which is independent from $\sfc$ vs
$\eta_m$. It further provides an additional consistency check
since this curve and $\sfc$ vs $\eta_m$ need to cross at the same
$\eta_m$ value as $\sfc_\mathrm{r}$ vs $\eta_m$ vanishes (namely
at $\eta_m = 1$).

If the assumptions on the composition of the samples leading to
the CEA are correct, then the specific heat jump at $\Tc$ is
proportional to the condensation energy:
\begin{equation}\label{eqn:DeltaCb}
  \left. \Delta \Cp \right \vert_{\Tc} = b \cdot \ce.
\end{equation}
Here, $b$ is a constant of proportionality thermodynamically
related to the critical field by Eq.~(\ref{eqn:dHc_dT}).

\subsection{Discussion}

As the analysis shows, the relevant parameter when measuring the
total mass of a sample experimentally is the superconducting mass
fraction $\eta_m$. All the relevant properties are linear in the
mass fraction and therefore in the condensation energy, which can
be readily extracted from specific heat measurements by
integration. The reason for this is that the total mass $m$ of the
sample is used for the conversion from extensive into intensive
form, not the volume or the a.o.s. In the case where another
sample property is measured, the analysis remains analogous, but
the parameters are linear in the corresponding superconducting
fraction instead of the \smf. These variations are discussed in
more detail in the Appendix.

We would like to mention, that though in principle still possible,
performing the CEA would be more difficult when dealing with a
nodal superconductor. In this case the superconducting heat
capacity goes to zero not exponentially, but following a
power-law. As a consequence, there is still some contribution from
the superconductor to $\sfc_\mathrm{r}$ at all but the very lowest
temperatures. The same argument holds for the analysis of the
slope of $\Delta \Cp$ vs $T$ (Eq.~\ref{eqn:DeltaC}). The CEA would
still work fine, but data to even lower temperatures might be
needed.

Plotting the data as in Fig.~\ref{fig:BruhwilerPlot} (here for
\fu) allows $\sfc_1$, $\sfc_2$, and $\ce_1$ to be extracted at a
glance. At the same time, the plot serves as a check for the
consistency of the analysis, since it is immediately apparent if
the lines cross each other and the $x$- and $y$-axes at the
correct points: $\sfc_\mathrm{r}$ needs to vanish at the same
$\eta_m$ value as $\sfc$ vs $\eta_m$ and the slope of $\Delta
\Cp(T)$ vs $\eta_m$ cross. Also, the $y$-intercepts of $\sfc$ vs
$\eta_m$ and $\sfc_\mathrm{r}$ vs $\eta_m$ need to be the same.

We would like to comment on the necessity to pinpoint $\eta_m=1$
for the extraction of the intrinsic parameters. In this study we
accomplish this by extrapolating $\sfc_\mathrm{r}$ to zero which
is possible because the second component is metallic and thus has
a non-vanishing $\sfc_2$. It is conceivable that $\eta_m=1$ could
also be pinpointed using some other technique. The ratio $\Delta
\Cp\vert_{T_\mathrm{c}}/(\sfc_1 \Tc)$ is a constant as a function
of $\eta_m$ if $\sfc_2 = 0$ and therefore it can be determined
using a single sample with an arbitrary volume fraction as long as
$\sfc_2 = 0$. $b$ is always independent of $\eta_m$ since it is
the proportionality factor between $\Delta
\Cp\vert_{T_\mathrm{c}}$ and $\ce$. If the true condensation
energy $\ce_1$ is to be extracted, we need to locate $\eta_m =1$
in any case. Since the normalized critical field slope $Q$
(defined in Tab.~\ref{tab:affected_param}) and $b$ are related
thermodynamically, also $Q$ can be determined independently of
$\eta_m=1$. In the case where $\sfc_2 = 0$, it follows that $\sfc
= \eta_m \sfc_1$ and thus $\sfc_1/\ce_1 = \sfc/\ce$ rendering
$1/(8\pi) \cdot (\sfc_1\Tc^2)/\ce_1$ independent of $\eta_m$. The
Sommerfeld coefficient $\sfc_1$ is only independent of $\eta_m$ if
$\sfc_1 = \sfc_2$ (Eq.~(\ref{eqn:gamma_b})). The reduced critical
field $h_\mathrm{c}$ and the deviation function $D(t)$ (defined
later in the text) are independent of $\eta_m$ since
$\ce_1(T)/\ce_1(0\unit{K}) = \ce(T)/\ce(0\unit{K})$. In summary,
$b$, $Q$, $h_\mathrm{c}$, and $D(t)$ can be determined using a
single sample regardless of its \smf. $\Delta
\Cp\vert_{T_\mathrm{c}}/(\sfc_1 \Tc)$ and $1/(8\pi) \cdot
(\sfc_1\Tc^2)/\ce_1$ can be extracted with a single sample if
$\sfc_2=0$ but several samples are needed if $\sfc_2 \ne 0$. For
$\ce_1(0\unit{K})$ and $\sfc_1$ it is imperative to measure many
samples with different \smf s and perform the CEA. There is only
one improbable exception: If $\sfc_1 = \sfc_2$, $\sfc_1$ is
independent of the \smf. These findings are listed in
Tab.~\ref{tab:affected_param}.

\begin{table}
\caption{\label{tab:affected_param} Compilation of thermodynamic
parameters for the extraction of which it is (a) necessary, (b)
necessary if $\sfc_2 \ne 0$, and (c) not necessary to locate the
ideal sample ($\eta_m=1$). When dealing with a metallic second
component, it is therefore essential to perform the condensation
energy analysis for the cases (a) and (b).}
\begin{ruledtabular}
\begin{tabular}{lr}
Parameter & Classification \\
\hline
$\ce_1(0\unit{K})$ & (a) \\
$\sfc_1$ & (a), if $\sfc_1 \ne \sfc_2$ \\
\hline
$\Delta \Cp\vert_{T_\mathrm{c}}/(\sfc_1 \Tc)$ & (b) \\
$\frac{1}{8\pi} \frac{\sfc_1\Tc^2}{\ce_1}$ & (b) \\
\hline
$b$ & (c) \\
$Q \equiv - \frac{2\Tc}{\Hc(0)} \left.\frac{\mathrm{d}\Hc}{\mathrm{d}T}\right\vert_{T_\mathrm{c}} $ & (c) \\
$h_\mathrm{c}(t)$, $D(t)$ & (c) \\
\end{tabular}
\end{ruledtabular}
\end{table}

Since the jump in the heat capacity $\left. \Delta \Cp \right
\vert_{\Tc}$ is proportional to $\ce$, thermodynamic parameters
may be analyzed in terms of $\Delta \Cp \vert_{\Tc}$ instead of
$\ce = \ce_1 \eta_m$. This may eliminate some scatter in the data
because $\left. \Delta \Cp \right \vert_{\Tc}$ is more directly
determined than $\ce$. Since in this case the condensation energy
$\ce_1$ remains unknown, however, many intrinsic parameters cannot
be extracted. This alternative approach is particularly helpful
when the heat capacity data do not extend to low enough
temperature to yield the condensation energy by integration,
nevertheless enabling the determination of the intrinsic $\sfc_1$,
$\sfc_2$, and $\left. \Delta \Cp \right \vert_{\Tc}$.

\section{Application to \fu}

Using varied conditions in the preparation procedure, \fu\ samples
with \svf s from $34\unit{\%}$ to $83\unit{\%}$ have been
prepared. The varying superconducting fraction, which is an
important prerequisite for the application of the CEA, is caused
by the presence of OsO$_2$ as determined by X-ray diffraction
analysis. We have therefore also measured a sample of the starting
material OsO$_2$ and included it in the data analysis.

The polycrystalline samples all show the transition to the bulk
superconducting state in resistivity and magnetization
measurements. They typically weigh about $10$ to $20 \unit{mg}$.
The samples, labeled A to F, have been synthesized by a procedure
originally described in Ref.~\cite{YoMuMaHi2004} and described in
more detail in Ref.~\cite{KaZhBrBaKa2004}. An overview over the
measured parameters is provided in Tab.~\ref{tab:samples}.

The specific heat was measured in a physical properties
measurement apparatus using an adiabatic relaxation technique
(Quantum Design, PPMS). Figure \ref{fig:CdivT_vs_T} shows the
specific heat $\Cp/T$ vs $T$ for two exemplary samples in the
superconducting and normal state ($0\unit{T}$ and $12\unit{T}$).
The residual Sommerfeld coefficient $\sfc_\mathrm{r}$, the
normal-state coefficient $\sfc$, and the specific heat jump at
$\Tc$, $\Delta \Cp \vert_{\Tc}/\Tc$, vary among the various
samples measured in a way that is a typical signature of a
metallic second component. These observations form the basis for
the condensation energy analysis (CEA) presented above. We
therefore analyze our data on \fu\ accordingly.

\begin{table}
\caption{\label{tab:samples} Overview over the parameters
extracted from the six samples measured in this study. The \svf s
$\eta_V$ range from $34\unit{\%}$ to $83\unit{\%}$ due to varied
conditions in the synthesis procedure.}
\begin{ruledtabular}
\begin{tabular}{lcccccc}

 & A & B & C & D & E & F \\
\hline

$\gamma$ ($\mu$J/g/K$^2$) & $35.5$ & $54.3$ & $60.7$ &$61.7$
& $58.8$ & $65.0$ \\

$\beta$ ($\mu$J/g/K$^{4}$) & $3.26$ & $1.38$ &
$2.23$ & $0.72$ & $1.22$ & $1.60$ \\

$\sfc_\mathrm{r}$ ($\mu$J/g/K$^2$) & $17.5$ & $14.8$ & $15.2$ &
$8.9$ & $7.7$ & $4.6$ \\

$-\frac{\mathrm{d}(\Delta \Cp)}{\mathrm{d}T}$ ($\mu$J/g/K$^2$) &
$21.7$ & $40.8$ & $47.8$ & $53.5$ & $52.5$ & $62.0$ \\

$\ce(0\unit{K})$ ($\mu$J/g) & $211$ & $378$ & $465$
& $528$ & $607$ & $644$ \\

$\Tc$ (K)\footnotemark[1] & $6.25$ & $6.22$ & $6.36$ &6.36
& 6.47 & 6.37 \\

Est.~svf $\frac{\sfc - \sfc_\mathrm{r}}{\sfc}$
(\%)\footnotemark[3] & $50.7$ & $72.7$ & $75.0$ &85.6
& 86.9 & 92.9 \\

Svf $\eta_V$ (\%)\footnotemark[2]\,\footnotemark[3] & $34.0$ &
$55.5$ & $65.1$ & $71.6$ & $79.2$ & $82.6$ \\

Rel.~error (\%)\footnotemark[2] & $49.0$ & $31.2$ & $15.1$
& $19.5$ & $9.7$ & $12.6$ \\

Smf $\eta_m$ (\%)\footnotemark[3] & $24.5$ & $44.0$ & $54.1$
& $61.4$ & $70.6$ & $74.9$ \\

\end{tabular}
\end{ruledtabular}
\footnotetext[1]{From $\Cp/T$ vs $T$ data
(Fig.~\ref{fig:CdivT_vs_T}).} \footnotetext[2]{With the second
component being OsO$_2$, $\mu \approx 0.63$.}
\footnotetext[3]{svf: \svf, smf: \smf.}
\end{table}

\subsection{Normal state properties and determination of the condensation energy
$\ce_1$}\label{sec:normal_props}

From the normal-state curves at $12\unit{T}$, we extract the
Sommerfeld coefficients $\gamma =
\lim_{T\to0\unit{K}}C_{\mathrm{p}}/T$ by fitting the data from
$1\unit{K}$ to $4\unit{K}$ to $C_{\mathrm{p}}/T = \gamma + \beta
T^2 + D T^4$. The $D$-coefficients obtained by this fit procedure
are essentially zero. Fits performed with restraining $D$ to zero
result in $\sfc$ values differing by only $0.6\unit{\%}$ or less
compared to the ones where $D$ is not restrained. The only
exception is sample A with a difference of $2.8\unit{\%}$, which
we attribute to an impurity concentration. We have used data down
to $1\unit{K}$ only for the fits because moderate contributions to
the heat capacity can be expected from the nucleus of Rb at the
lowest temperatures \footnote{We write the contribution from the
Rb nucleus in the usual high temperature expanded form
$C_\mathrm{nuc} = \eta_m A (\mu_0 H / T)^2$, where $\eta_m$
accounts for the non-ideal samples and $A = 3.54 \unit{nJ/g \cdot
K/T^2}$. At $\mu_0 H = 12\unit{T}$ the nuclear heat capacity
amounts to a negligible $C_\mathrm{nuc}/(\sfc_1 T) \approx
0.6\unit{\%}$ at $1\unit{K}$ and a moderate $7\unit{\%}$ at
$0.45\unit{K}$.}. The upper temperature limit of $4\unit{K}$ is
chosen such as to avoid any influence of possible low-lying
non-Debye like modes, which cannot be parameterized by
$C_{\mathrm{p}}/T = \gamma + \beta T^2$.

\begin{figure}
  \includegraphics{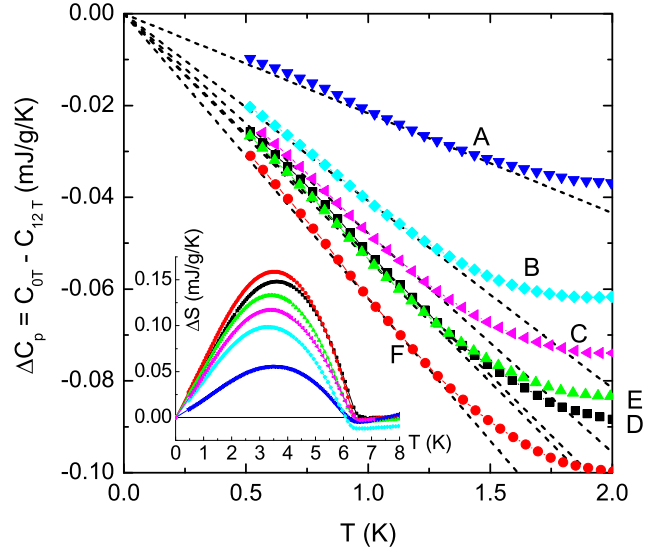}
  \caption{\label{fig:DeltaC_vs_T} (Color online) Low temperature
  heat capacity difference $\Delta \Cp$ for the six samples measured
  in this study. At these temperatures
  $C_\mathrm{es} \ll \sfc_1 T$ and therefore $\Delta
\Cp = - \sfc_1 \eta_m T$, i.e.~$\Delta \Cp$ is linear in $T$ and
the slope is proportional to $\eta_m$, varying from sample to
sample. The inset shows the difference in entropy
 between the superconducting and normal state
 $\Delta S (T) = \int_{T}^{\Tc} \Delta \Cp(T')/T' \mathrm{d}T'$.}
\end{figure}

\begin{figure}
  \includegraphics{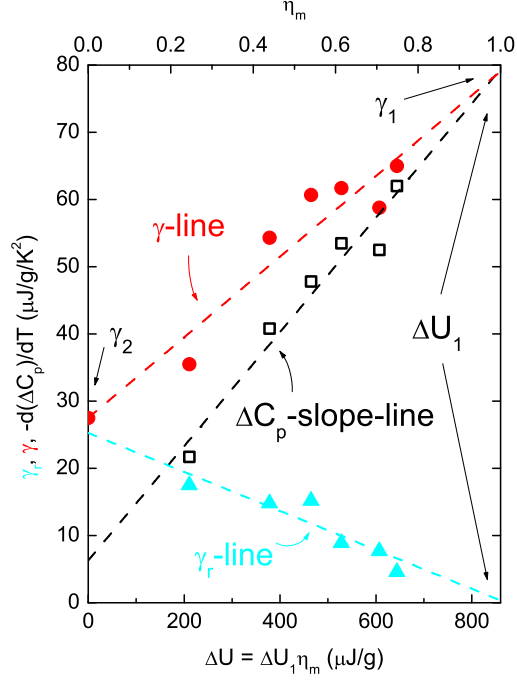}
  \caption{\label{fig:BruhwilerPlot} (Color online) Summary of the
  data analysis for the six different \fu\ samples and the OsO$_2$
  sample at $\ce = 0$. Plotted vs the
  condensation energy are the measured Sommerfeld coefficient
  $\sfc$, the residual Sommerfeld coefficient
$\sfc_\mathrm{r}$, and the slope from $\Delta \Cp$ vs $T$ at low
temperatures. All quantities are directly extracted from
measurement (Figs.~\ref{fig:CdivT_vs_T} and \ref{fig:DeltaC_vs_T})
and they all depend linearly on the \smf\ $\eta_m = \ce/\ce_1$,
see Eqs.~(\ref{eqn:gamma_res_b}), (\ref{eqn:gamma_b}), and
(\ref{eqn:DeltaC}). Thus a linear fit of these data is shown as
$\sfc$-line, $\sfc_\mathrm{r}$-line, and $\Delta \Cp$-slope-line
respectively. The plot allows the extraction of $\sfc_1$,
$\sfc_2$, and $\ce_1$ at a glance and serves as a check for the
consistency of the analysis.}
\end{figure}

The residual coefficient $\sfc_\mathrm{r}$ is the $\Cp/T$ value at
$1\unit{K}$ and $0\unit{T}$ and $\Tc$ is chosen such as to balance
the entropy (refer to Fig.~\ref{fig:CdivT_vs_T}). We further
extract the negative of the slope of $\Delta \Cp$ vs $T$, which at
low temperatures is proportional to $\eta_m$
(Eq.~(\ref{eqn:DeltaC}), Fig.~\ref{fig:DeltaC_vs_T}). The values
obtained for the various samples are summarized in
Tab.~\ref{tab:samples}.

The actual integration for the entropy difference $\Delta S$
according to Eq.~(\ref{eqn:DeltaS_of_T}) is performed from
$0\unit{K}$ to $T$ using the relation $\int_T^{\Tc} =
\int_{0\unit{K}}^{\Tc} - \int_{0\unit{K}}^T$ and assuming balanced
entropy, i.e.~$\int_{0\unit{K}}^{\Tc}\Delta \Cp / T' \mathrm{d}T'
= 0$. To integrate starting from $0\unit{K}$, the $\Cp/T$ curves
have to be extrapolated to $0\unit{K}$ from their lowest value at
$0.5\unit{K}$. For this purpose we use the fitted Sommerfeld
coefficients $\sfc$ (s.~Tab.~\ref{tab:samples}) for the $\Cp/T$
value at $0\unit{K}$ for the $12\unit{T}$ curves. Similarly, we
use the $\Cp/T$ value at the lowest temperature of the $0\unit{T}$
curve and use it to extrapolate a constant $\Cp/T$ to $0\unit{K}$.
The resulting $\Delta S$ curves are shown in the inset of
Fig.~\ref{fig:DeltaC_vs_T} and the condensation energies
$\ce(0\unit{K})$ are listed in Tab.~\ref{tab:samples}. The
$\ce(T)$ curves derived by integrating $\Delta S(T)$
(Eq.~(\ref{eqn:Hc_of_T})) were also obtained using the relation
$\int_T^{\Tc} = \int_{0\unit{K}}^{\Tc} - \int_{0\unit{K}}^T$
together with $\int_{0\unit{K}}^{\Tc} \Delta S \mathrm{d}T' =
\ce(0\unit{K})$.

The essential steps of the CEA are shown in
Fig.~\ref{fig:BruhwilerPlot}: The measured Sommerfeld coefficient
$\sfc$, the residual Sommerfeld coefficient $\sfc_\mathrm{r}$, and
the slope from $\Delta \Cp$ vs $T$ at low temperatures are plotted
as a function of the condensation energy $\ce$. All these
quantities are directly extracted from measurement and they all
depend linearly on the \smf\ $\eta_m = \ce/\ce_1$, see
Eqs.~(\ref{eqn:gamma_res_b}), (\ref{eqn:gamma_b}), and
(\ref{eqn:DeltaC}). We therefore perform a linear fit of these
data calling the fitted lines $\sfc$-line, $\sfc_\mathrm{r}$-line,
and $\Delta \Cp$-slope-line respectively, and use them to extract
parameters and perform consistency checks. For the fit of the
$\sfc$-line only we have included the data point resulting from
the OsO$_2$ sample ($\sfc_2 = 27.5\unit{\mu J/g/K^2}$ at $\ce=0$).

For the fit of the $\Delta \Cp$-slope-line we have refrained from
restraining the $y$-intercept to $0$, even though it would be
justified from Eq.~(\ref{eqn:DeltaC}). This results in a small,
finite intercept which we attribute to a minute $C_\mathrm{mag}$
rather than systematic error in $\ce$, since $\Delta \Cp$ vs $\ce$
shows no similar offset. This explanation is also suggested by the
heat capacity data at $12\unit{T}$ and $0\unit{T}$, which slightly
differ above $\Tc$. This minute refinement will be further
discussed later in the context of $C_\mathrm{es}$ and in Appendix
\ref{sec:Incl_Cmag}.

The fully superconducting sample, i.e.~the sample with $\eta_m =
\eta_V = 1$, is defined by the $x$-intercept of the
$\sfc_\mathrm{r}$-line. This defines the value of
$\ce\vert_{\eta_m=1}$ and with that it defines the condensation
energy $\ce_1$ of the superconductor. The value obtained in this
way for $\ce_1$ is $872\unit{\mu J/g}$. The $y$-intercept of the
$\sfc_\mathrm{r}$-line of $(25.3 \pm 3)\unit{\mu J/g/K^2}$ agrees
well with the measured Sommerfeld coefficient $\sfc_2$ of OsO$_2$
of $27.5\unit{\mu J/g/K^2}$, providing a self-consistency check.

The Sommerfeld coefficient of component one, $\sfc_1$, is the
ordinate at the point where the $\sfc$-line and the $\Delta
\Cp$-slope-line cross. According to the CEA, this happens at
$\eta_m = 1$, i.e.~at the same $\eta_m$ value as $\sfc_\mathrm{r}$
vanishes. The crossing point thus is another means to find
$\eta_m=1$. The point of intersection of the two lines is at
$849\unit{\mu J/g}$, in good agreement with the $872\unit{\mu
J/g}$ obtained from the $\sfc_\mathrm{r}$-line. As the effective
$\ce_1$ value we thus use $860\unit{\mu J/g}$
($483\unit{mJ/mol_\mathrm{f.u.}}$), yielding a Sommerfeld
coefficient for \fu\ of $\sfc_1 = 79\unit{\mu J/g/K^2}$
($44\unit{mJ/mol_\mathrm{f.u.}/K^2}$).

In this study, we have performed separate linear fits for the
$\sfc$-line, $\sfc_\mathrm{r}$-line, and $\Delta \Cp$-slope-line
respectively. Good agreement among the $\sfc_2$ and $\ce_1$ values
obtained with independent lines confirm that our assumptions on
the composition of the samples (motivated by the X-ray
diffractograms) leading to the CEA are correct. For further
refinement of the parameters, it would thus be appropriate to
perform a simultaneous fit of all three lines with constraints on
the relevant intercepts.

\begin{table}
\caption{\label{tab:TD_params} Thermodynamic parameters of \fu. }
\begin{ruledtabular}
\begin{tabular}{lr}
Parameter & Value \\
\hline
$\xi(0\unit{K})$ & $74\unit{\AA}$ \\
$\lambda_\mathrm{eff}(0\unit{K})$ & $252\unit{nm}$ \\
$\kappa(\Tc)$, $\kappa(0\unit{K})$ & $23$, $34$ \\
\hline
$\Delta \Cp\vert_{T_\mathrm{c}}/(\sfc_1 \Tc)$ & 1.9 \\
$\lambda_\mathrm{ep}$ & $1\pm0.1$ \\
$b$ & $(1.12\pm0.03)\unit{K^{-1}}$ \footnotemark[1] \\
$b/\Tc$ & $(0.175\pm0.005)\unit{K^{-2}}$ \footnotemark[1] \\
$\ce_1(0\unit{K})$ & $860\unit{\mu J/g}$ ($483\unit{mJ/mol_\mathrm{f.u.}}$) \\
\hline
$\Hc(0\unit{K})$ & $1249\unit{Oe}$ \\
$H_\mathrm{c1}(0\unit{K})$ & $92\unit{Oe}$ \\
$H_\mathrm{c2}(0\unit{K})$ & $60000\unit{Oe}$ \footnotemark[2] \\
$-\mathrm{d}\Hc/\mathrm{d}T\vert_{T_\mathrm{c}}$ & $369 \unit{Oe/K}$ \\
$-\mathrm{d}H_\mathrm{c2}/\mathrm{d}T\vert_{T_\mathrm{c}}$ & $12000 \unit{Oe/K}$ \footnotemark[2] \\
$Q \equiv - \frac{2\Tc}{\Hc(0)} \left.\frac{\mathrm{d}\Hc}{\mathrm{d}T}\right\vert_{T_\mathrm{c}} $ & $3.79\pm0.05$ \\
$k_\mathrm{B}\Tc/(\hbar\omega_\mathrm{ln})$ & $0.06$ \\
$\frac{2\Delta(0\unit{K})}{k_\mathrm{B}\Tc}$ & $3.87$\\
$\frac{1}{8\pi} \frac{\sfc_1\Tc^2}{\ce_1}$ & $0.15$\\
\hline
$\sfc_1$ & $79\unit{\mu J/g/K^2}$ ($44\unit{mJ/mol_\mathrm{f.u.}/K^2}$) \\
\end{tabular}
\end{ruledtabular}
\footnotetext[1]{$b \cdot \Tc = Q^2/2$ has a universal value of
$5.99$ for a weak coupling BCS superconductor. This results in
$b/\Tc = 0.146\unit{K^{-2}}$ and $b = 0.94\unit{K^{-1}}$ for a BCS
superconductor with $\Tc=6.4\unit{K}$.}\footnotetext[2]{From
Ref.~\cite{BrKaZhKaBa2004}}
\end{table}

\subsection{Superconducting properties and density of states enhancement}\label{sec:SC_props}

\begin{figure}
  \includegraphics{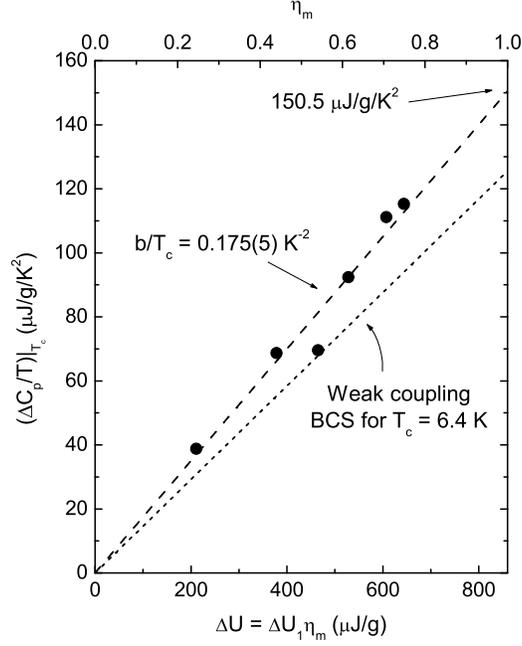}
  \caption{\label{fig:DC_vs_S} The measured specific heat jump at
   $\Tc$, $\Delta \Cp \vert_{\Tc}/\Tc$, is
proportional to the measured condensation energy. A linear fit
using $\left. \Delta \Cp \right \vert_{\Tc} / \Tc = b/\Tc \cdot
\ce$ with $b/\Tc$ as a fit parameter gives $b/\Tc =
(0.175\pm0.005)\unit{K^{-2}}$. For comparison, the BCS weak
coupling $b/\Tc = 0.146$ for $\Tc = 6.4 \unit{K}$ is also shown.
The specific heat jump for \fu\ is $150.5\unit{\mu J/g/K^2}$,
resulting in $\Delta \Cp\vert_{T_\mathrm{c}}/(\sfc_1 \Tc) = 1.9$,
and thus a coupling parameter $\lambda_\mathrm{ep} \approx 1$.}
\end{figure}

The specific heat jump at $\Tc$, $\Delta \Cp \vert_{\Tc}/\Tc$, is
extracted from the $\Cp/T$ vs T plot in Fig.~\ref{fig:CdivT_vs_T}
and plotted vs $\ce$ in Fig.~\ref{fig:DC_vs_S}. As expected,
$\Delta \Cp \vert_{\Tc}/\Tc$ is proportional to the condensation
energy $\ce$. A linear fit using $\left. \Delta \Cp \right
\vert_{\Tc} / \Tc = b/\Tc \cdot \ce$ with $b/\Tc$ as a fit
parameter gives $b/\Tc = (0.175\pm0.005)\unit{K^{-2}}$, and a
specific heat jump for the pure superconductor of $\Delta \Cp
\vert_{\Tc}/\Tc = 150.5\unit{\mu J/g/K^2}$. This results in a
normalized specific heat anomaly $\Delta
\Cp\vert_{T_\mathrm{c}}/(\sfc_1 \Tc) = 1.9$. The accuracy of this
value depends mainly on the accuracy of the condensation energy
$\ce_1$, since the error in $b/\Tc$ is rather small. Since both
$\sfc$ and $\left. \Delta \Cp \right \vert_{\Tc} / \Tc$ are linear
in $\ce$ with a positive slope (in the case of \fu), the ratio
$\Delta \Cp\vert_{T_\mathrm{c}}/(\sfc_1 \Tc)$ is relatively stable
against an error in $\ce$.

The normalized specific heat jump $\Delta
\Cp\vert_{T_\mathrm{c}}/(\sfc_1 \Tc) = 1.9$ is significantly
larger than that for the weak-coupling case ($1.43$). It
corresponds to an electron-phonon coupling constant
$\lambda_\mathrm{ep} = 2 \int_0^\infty \alpha^2 F(\omega)/\omega
\, \mathrm{d}\omega \approx 1$ \cite{MaCoCa1987}, i.e.~\fu\ is a
superconductor in the intermediate-coupling regime. Here,
$\alpha^2 F(\omega)$ is the electron-phonon spectral density.
Using the calculated band Sommerfeld coefficient $\sfc_\mathrm{b}
= 17.8\unit{\mu J/g/K^2}$ of \fu\ from Ref.~\cite{KuJePi2004}, the
present result indicates a significant enhancement of the
electronic specific heat of
$(1+\lambda_\mathrm{ep}+\lambda_\mathrm{add}) = (79.1\unit{\mu
J/g/K^2})/(17.8\unit{\mu J/g/K^2}) \approx 4.4$. This enhancement
surpasses the one found in Sr$_2$RuO$_4$ of about $3.8$ by
$16\unit{\%}$ \cite{SaMeYeShFr2004}. Additional to the
electron-phonon enhancement $\lambda_\mathrm{ep}$, there is a
strong enhancement of unknown origin $\lambda_\mathrm{add} \approx
2.4$. We use the band structure $\sfc_\mathrm{b}$ from another
Reference to estimate the uncertainty in this additional
enhancement: \citeauthor{SaMeYeShFr2004}
(Ref.~\cite{SaMeYeShFr2004}) have calculated the band Sommerfeld
coefficient for KOs$_2$O$_6$, which is $18\unit{\%}$ higher than
the one calculated in Ref.~\cite{KuJePi2004}. Assuming these
$18\unit{\%}$ to be the uncertainty in $\sfc_\mathrm{b}$ for \fu,
this would result in a $\lambda_\mathrm{add} \approx 2.1\pm0.3$.
In view of a calculated Stoner enhancement of the magnetic
susceptibility of roughly $2$ \cite{KuJePi2004}, we speculate that
the additional enhancement is due to spin correlation effects.

From $\ce_1(0\unit{K})$ we calculate a thermodynamic critical
field $\Hc(0\unit{K}) = 1249\unit{Oe}$. Together with the exact
thermodynamic relationship
\begin{equation}\label{eqn:dHc_dT}
  \left.\frac{\mathrm{d}\Hc}{\mathrm{d}T} \right \vert_{T_\mathrm{c}} = \frac{-1}{\sqrt{2}} \sqrt{b/\Tc} \Hc(0\unit{K})
\end{equation}
this gives a critical field slope of
$-\mathrm{d}\Hc/\mathrm{d}T\vert_{T_\mathrm{c}} = 369
\unit{Oe/K}$. The same relationship gives the normalized critical
field slope (also called "$Q$" in literature) $Q \equiv -
\frac{2\Tc}{\Hc(0)}
\left.\frac{\mathrm{d}\Hc}{\mathrm{d}T}\right\vert_{T_\mathrm{c}}
= \sqrt{2b\Tc} = 3.79\pm0.05$. To calculate the Ginzburg-Landau
parameter $\kappa$ at $\Tc$, we use
\begin{equation}\label{eqn:kappa}
  \kappa(\Tc) = \frac{1}{\sqrt{2}} \frac{ \left.\frac{\mathrm{d}H_\mathrm{c2}}{\mathrm{d}T} \right\vert_{T_\mathrm{c}} }
  { \left.\frac{\mathrm{d}\Hc}{\mathrm{d}T} \right\vert_{T_\mathrm{c}}} = \frac{ -\left.\frac{\mathrm{d}H_\mathrm{c2}}{\mathrm{d}T} \right\vert_{T_\mathrm{c}} }
  { \sqrt{b/\Tc} \Hc(0\unit{K})}
\end{equation}
to get $\kappa(\Tc) = 23$. For this we have used
$-\mathrm{d}H_\mathrm{c2}/\mathrm{d}T\vert_{T_\mathrm{c}} = 12000
\unit{Oe/K}$ from Ref.~\cite{BrKaZhKaBa2004}. At $0\unit{K}$ we
use $\kappa(0\unit{K}) = 1/\sqrt{2} \cdot
H_\mathrm{c2}(0\unit{K})/\Hc(0\unit{K})$ to get $\kappa(0\unit{K})
= 34$ using a reasonably extrapolated $H_\mathrm{c2}(0\unit{K}) =
60000\unit{Oe}$ \cite{BrKaZhKaBa2004}. Using
\begin{equation}\label{eqn:lambda_eff}
  \lambda_\mathrm{eff} = \sqrt{\frac{\kappa
  \Phi_0}{2\sqrt{2}\pi\mu_0\Hc}},
\end{equation}
where $\Phi_0$ is the magnetic flux quantum, we get the
penetration depth $\lambda_\mathrm{eff}(0\unit{K})= 252\unit{nm}$.
This is fairly close to the estimated $230\unit{nm}$ obtained from
magnetization measurements in
Ref.~\cite{KhEsKaKaZhBrGaDiShMaMaBaKe2004} and somewhat larger
than the $212\unit{nm}$ obtained from an analysis in
Ref.~\cite{SchKhKe2004}. The Ginzburg-Landau coherence length
results in $\xi(0\unit{K}) =
\lambda_\mathrm{eff}(0\unit{K})/\kappa(0\unit{K}) \approx
74\unit{\AA}$. Finally, the lower critical field results from
\begin{equation}\label{eqn:Hc1}
  H_\mathrm{c1} = \frac{\ln\kappa}{\sqrt{2}\kappa} \Hc
\end{equation}
and gives $H_\mathrm{c1}(0\unit{K}) = 92\unit{Oe}$.

Using the approximate semiphenomenological forms of the
strong-coupling correction to the weak coupling BCS ratios which
hold for many superconductors \cite{MaCa1986},
\begin{equation}\label{eqn:DeltaC_of_TcdivOm}
  \frac{\Delta \Cp\vert_{T_\mathrm{c}}} {\sfc_1 \Tc} = 1.43 \left [ 1 +
  53 \left ( \frac{k_\mathrm{B}\Tc}{\hbar\omega_\mathrm{ln}} \right )^2 \ln \left ( \frac{\hbar\omega_\mathrm{ln}}{3k_\mathrm{B}\Tc} \right
  ) \right ]
\end{equation}
and
\begin{equation}\label{eqn:gap_of_TcdivOm}
  \frac{2\Delta (0\unit{K})} {k_\mathrm{B} \Tc} = 3.53 \left [ 1 +
  12.5 \left ( \frac{k_\mathrm{B}\Tc}{\hbar\omega_\mathrm{ln}} \right )^2 \ln \left ( \frac{\hbar\omega_\mathrm{ln}}{2k_\mathrm{B}\Tc} \right
  ) \right ],
\end{equation}
we get a strong-coupling parameter $k_\mathrm{B}\Tc/(\hbar
\omega_\mathrm{ln}) = 0.06$ and $2\Delta(0\unit{K}) /
(k_\mathrm{B}\Tc) = 3.87$. Here $\hbar \omega_\mathrm{ln}$ is the
Allen-Dynes expression for the average phonon energy. The
Sommerfeld coefficient, the condensation energy, and the critical
temperature result in  $1/(8\pi) \cdot (\sfc_1\Tc^2)/\ce_1 =
0.15$, a value well in the range of conventional phonon-mediated
superconductors \cite{MaCa1986}.

The values discussed in the previous sections are summarized in
Tab.~\ref{tab:TD_params}.

We extract the superconducting electronic specific heat
$C_\mathrm{es}(T)$ in the way described according to
Eq.~(\ref{eqn:Ces}). To correct for the metallic second component
present, the specific heat difference $\Delta \Cp$ needs to be
scaled with the inverse \smf\ and the added electronic specific
heat from the normal state, $C_\mathrm{en}(T) = \sfc_1 T$, has to
be calculated using the Sommerfeld coefficient $\sfc_1$ of the
pure superconductor. As Figure \ref{fig:Ces_vs_TdivTc} shows, the
electronic superconducting specific heat for various samples
indeed shows a universal behavior when scaled appropriately.

The data at very low temperatures are susceptible to even a minute
$C_\mathrm{mag}$ term present in the $12\unit{T}$ heat capacity
(c.f.~Eq.~(\ref{eqn:C_12T_0T})). For a logarithmic plot of
$C_\mathrm{es}$, where the behavior at low temperatures is much
expanded, it is therefore necessary to further fine-tune the CEA
and account for such a small additional term. A contribution
consistent with the deviation of the $y$-intercept from $0$ of the
$\Delta \Cp$-slope-line (Fig.~\ref{fig:BruhwilerPlot}) amounts to
a few percent of the normal state specific heat $\sfc_1 T$. The
refined analysis results in a superconducting electronic specific
heat for \fu\ shown in the inset of Fig.~\ref{fig:Ces_vs_TdivTc}
on a logarithmic scale. The details of this refined analysis are
elaborated later in Appendix \ref{sec:Incl_Cmag}. It decreases in
close quantitative agreement with conventional superconducting
behavior. The solid line is a best fit to the data and indicates
the expected behavior from an isotropic gap: $C_\mathrm{es} = 9 \,
\sfc_1 \, T_\mathrm{c} \exp(-1.55 \, T_\mathrm{c}/T)$. The
dash-dotted line from weak coupling BCS also assuming an isotropic
gap is shown for comparison: For $2.5<T_\mathrm{c}/T<6$, the
specific heat approximately follows an exponential behavior
$C_\mathrm{es} = 8.5 \, \gamma \, T_\mathrm{c} \exp(-1.44 \,
T_\mathrm{c}/T)$ \cite{BaSch1961}. At low temperatures, the
subtraction provides unreliable results, since the difference
becomes exceedingly small. Furthermore, scattering from various
minute impurities may play a role at such low temperatures, and it
is thus difficult to draw conclusions from the observed deviation.

\begin{figure}
  \includegraphics{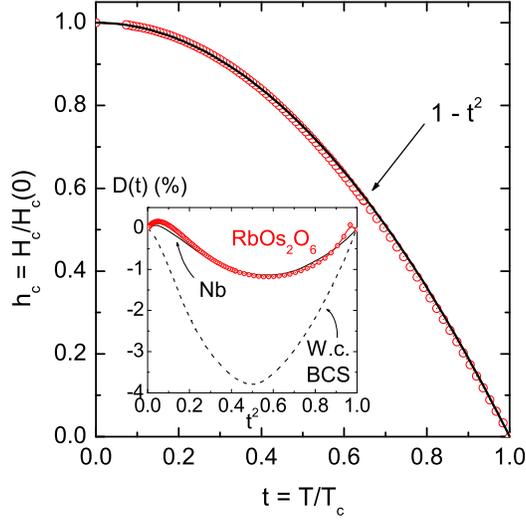}
  \caption{\label{fig:hc_vs_t} (Color online) The reduced critical field $h_\mathrm{c}(t) :=
H_\mathrm{c}(T)/H_\mathrm{c}(0\unit{K}) = \sqrt{\ce_1(T)/\ce_1
(0\unit{K})}$ of sample F. The inset shows the deviation $D(t) :=
h_\mathrm{c}(t) - (1-t^2)$ from a simple $1-t^2$ behavior for a
weak coupling BCS superconductor, niobium, and \fu. The deviation
function of \fu\ closely matches the data for Nb from
Ref.~\cite{LeBo1964}. The critical field at $0\unit{K}$ is
$\Hc(0\unit{K}) = 1249\unit{Oe}$.}
\end{figure}

We calculate the reduced critical field $h_\mathrm{c}(t) :=
H_\mathrm{c}(T)/H_\mathrm{c}(0\unit{K}) = \sqrt{\ce_1(T)/\ce_1
(0\unit{K})}$ which is independent of the \smf\ of a sample,
because $\eta_m$ cancels when taking the ratio $\ce_1(T)/\ce_1
(0\unit{K})$ (Tab.~\ref{tab:affected_param}). The deviation
function $D(t) := h_\mathrm{c}(t) - (1-t^2)$, where $t:=T/\Tc$,
measures the deviation of the critical field from a simple $1-t^2$
behavior. Figure \ref{fig:hc_vs_t} shows both these quantities as
a function of temperature for sample F. The deviation function of
\fu\ closely matches that of Nb from Ref.~\cite{LeBo1964}, which
is in line with the observation that the coupling constant
$\lambda_\mathrm{ep} \approx 1$ is the same as that of Nb, within
experimental uncertainty.

\section{Conclusion}

A detailed analysis is presented for the extraction of bulk
thermodynamic parameters for samples containing a superconductor
of interest and a second metallic non-superconducting component.
Since parameter \textit{estimates} can be significantly off even
for large superconducting fractions, the analysis is essential for
the extraction of intrinsic properties. We have developed all the
formulas necessary for the analysis and point to important
consistency checks. We apply the analysis, which we call
condensation energy analysis (CEA), to \fu\ to extract its
intrinsic thermodynamic parameters. The main results are as
follows: \fu\ is a strong type-II superconductor
($\kappa(\Tc)=23$) in the intermediate-coupling regime comparable
to niobium ($\lambda_\mathrm{ep} \approx 1$). From the upper
critical field $\mu_0 H_{\mathrm{c2}} \approx 6\unit{T}$ at
$0\unit{K}$, we estimate a Ginzburg-Landau coherence length $\xi
\approx 74\unit{\AA}$. The condensation energy is $860\unit{\mu
J/g}$ ($483\unit{mJ/mol_\mathrm{f.u.}}$) resulting in $1/(8\pi)
\cdot (\sfc_1\Tc^2)/\ce_1 \approx 0.15$, a value well in the range
of conventional phonon-mediated superconductors. The
superconducting electronic specific heat indicates conventional
s-wave pairing. \fu\ has a high Sommerfeld coefficient for a
pyrochlore of $79\unit{\mu J/g/K^2}$
($44\unit{mJ/mol_\mathrm{f.u.}/K^2}$) and thus a remarkably large
enhancement over the calculated band coefficient of $3.8$ to
$4.4$. We speculate that the origin of this mass enhancement lies
in the $3$-dimensional triangular nature of the pyrochlore
lattice.

This study was partly supported by the Swiss National Science
Foundation.

\newpage

\appendix

\section{Approximations}

To perform the CEA, more than one sample needs to be measured with
the samples having various fractions of component one and two. If
only the data from one sample is available, it is possible to
approximate the Sommerfeld coefficient and the \svf\ to some
extent. The validity of the approximations depends on the
properties of the material under investigation.

\subsection{Sommerfeld coefficient}

The true Sommerfeld coefficient of component one, $\sfc_1$, may be
approximated by dividing the Sommerfeld coefficient $\sfc$ by the
estimated superconducting volume fraction
$(\sfc-\sfc_\mathrm{r})/\sfc$ (see below):
\begin{equation}\label{eqn:gamma_approx}
  \frac{\sfc^2}{\sfc - \sfc_\mathrm{r}} = \sfc_1 + \delta (2 \sfc_2 -
  \sfc_1) + \delta^2 \frac{\sfc_2^2}{(1-\delta)\sfc_1}
  \stackrel{\delta \rightarrow 0}{\rightarrow} \sfc_1,
\end{equation}
where $\delta := 1-\eta_m$ is the deviation from the ideal sample.
This approximation is easily calculated from the measured
coefficients $\sfc$ and $\sfc_\mathrm{r}$ and also makes sense
intuitively. A criterion ensuring that $\sfc^2/(\sfc -
\sfc_\mathrm{r})$ is a better approximation to $\sfc_1$ than
$\sfc$ itself is
\begin{equation}\label{eqn:gamma_approx_criterion}
  \left \vert \left. \frac{\mathrm{d}}{\mathrm{d}\eta_m} \right \vert_{\eta_m=1} \left (\frac{\sfc^2}{\sfc - \sfc_\mathrm{r}} \right) \right \vert
  = \vert \sfc_1 - 2 \sfc_2 \vert < \vert \sfc_1 - \sfc_2 \vert =
  \left \vert \left. \frac{\mathrm{d}\sfc}{\mathrm{d}\eta_m} \right \vert_{\eta_m=1} \right
  \vert.
\end{equation}
This criterion tests the rate of departure of the approximation
$\sfc^2/(\sfc - \sfc_\mathrm{r})$ and the measured Sommerfeld
coefficient $\sfc$ from the true Sommerfeld coefficient $\sfc_1$
at $\eta_m = 1$. Strictly speaking, the criterion is thus only
valid locally around $\eta_m = 1$. Depending on the material
details, $\sfc^2/(\sfc - \sfc_\mathrm{r})$ or $\sfc$ may be the
better approximation at smaller volume fractions or they may both
be useless altogether. At very small volume fractions,
$\sfc^2/(\sfc - \sfc_\mathrm{r})$ is always a bad approximation,
since it diverges like $1/\eta_m$.

\begin{figure}
  \includegraphics{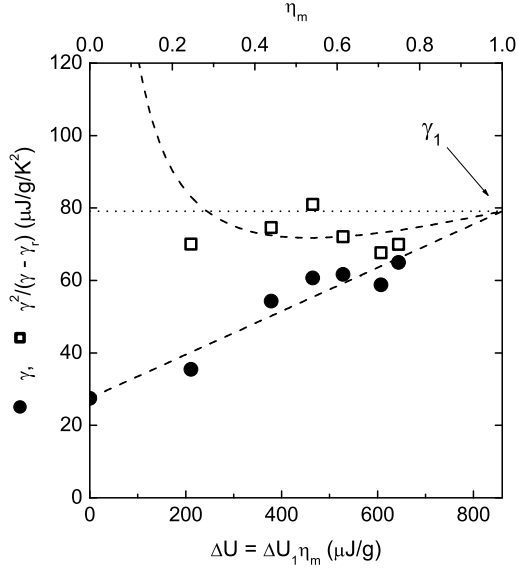}
  \caption{\label{fig:gamma_vs_DeltaU} Two different approaches to
  estimate the Sommerfeld coefficient of \fu: The estimate
  $\sfc^2/(\sfc-\sfc_\mathrm{r})$ compared to the coefficient
  $\sfc$. Down to a superconducting mass fraction of
$\eta_m \approx 20\unit{\%}$, $\sfc^2/(\sfc-\sfc_\mathrm{r})$ is
the better approximation to the true Sommerfeld coefficient
$\sfc_1$ than $\sfc$.}
\end{figure}

How does the estimated Sommerfeld coefficient
$\sfc^2/(\sfc-\sfc_\mathrm{r})$ compare to the measured
coefficient $\sfc$ in the case of \fu? Using the above determined
values for $\sfc_2$ and $\sfc_1$ we find that the criterion
Eq.~(\ref{eqn:gamma_approx_criterion}) is fulfilled: $\vert \sfc_1
- 2 \sfc_2 \vert = 24.1 \unit{\mu J/g/K^2} < 51.6 \unit{\mu
J/g/K^2} = \vert \sfc_1 - \sfc_2 \vert$. At high \smf s,
$\sfc^2/(\sfc-\sfc_\mathrm{r})$ is therefore a better
approximation to $\sfc_1$ than $\sfc$.
Fig.~\ref{fig:gamma_vs_DeltaU} shows that the estimate is
reasonable down to about $20\unit{\%}$ \smf\ in \fu.

\subsection{Superconducting volume fraction}

The true superconducting volume fraction is related to the \smf\
by
\begin{equation}\label{eqn:true_vol_frac}
  \eta_V = \frac{m_1/\rho_1}{m_1/\rho_1 + m_2/\rho_2} = \frac{\eta_m}{(1-\mu)\eta_m + \mu
  }.
\end{equation}
A possible approximation to $\eta_V$ easily obtained from
Eqs.~(\ref{eqn:gamma_res_b}) and (\ref{eqn:gamma_b}) is the ratio
$(\sfc-\sfc_\mathrm{r})/\sfc$ which is often used as an estimate:
\begin{equation}\label{eqn:estimated_vol_frac}
  \frac{\sfc-\sfc_\mathrm{r}}{\sfc} = \frac{\eta_m}{(1-\epsilon)\eta_m +
  \epsilon}.
\end{equation}
This ratio is easily extracted from specific heat measurements and
is therefore a popular way to extract the \svf. It has the same
functional dependence on $\eta_m$ as the true \svf, but it uses
the parameter $\epsilon$ instead of $\mu$. Depending on the
materials under investigation, this estimate can therefore be
significantly off: Comparing the estimate to the true
superconducting volume fraction $\eta_V$, we get
\begin{equation}\label{eqn:vol_frac_error}
  \frac{\frac{\sfc-\sfc_\mathrm{r}}{\sfc} - \eta_V}{\eta_V} =
  \frac{(\epsilon-\mu)(\eta_m-1)}{\eta_m-\epsilon(\eta_m-1)}.
\end{equation}
The relative error therefore is at maximum when $\eta_m=0$ and
amounts to $(\mu - \epsilon)/\epsilon$. It can be very large
depending on the mass densities, and the Sommerfeld coefficients
of the components. Even at a large superconducting fraction, the
error can be quite significant.

\begin{figure}
  \includegraphics{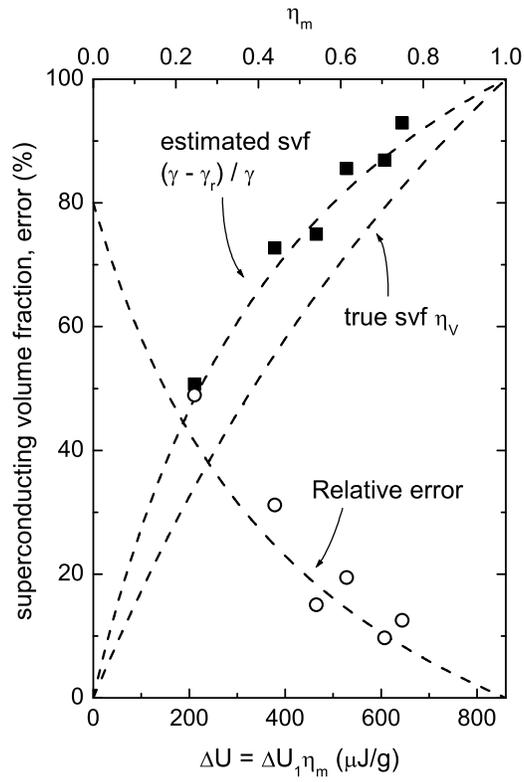}
  \caption{\label{fig:volfrac_vs_DeltaU} The estimated
   \svf\ (svf) $(\sfc-\sfc_\mathrm{r})/\sfc$ compared
to the true \svf\ $\eta_V$ plotted as a function of the
condensation energy $\ce$. With OsO$_2$ as the second component,
we calculated the ratio of the mass densities $\mu \approx 0.63$
and use it to evaluate $\eta_V$. The Figure also shows the
relative error which reaches a maximum of $80\unit{\%}$ at
$\eta_m=0$.}
\end{figure}

How does the estimated \svf\ $(\sfc-\sfc_\mathrm{r})/\sfc$ compare
to the true \svf\ $\eta_V$ in the case of \fu? In
Fig.~\ref{fig:volfrac_vs_DeltaU} we plot the true \svf\ and the
estimated fraction as a function of the condensation energy $\ce$.
With OsO$_2$ as the second component, we calculated the parameter
$\mu = \rho_1/\rho_2 \approx
(7.215\unit{g/cm^3})/(11.445\unit{g/cm^3}) \approx 0.63$ and use
it to evaluate $\eta_V$. For the estimated fraction we have used
the above values for $\sfc_2$ and $\sfc_1$ yielding $\epsilon
\approx 0.35$. The Figure also shows the relative error
(Eq.~(\ref{eqn:vol_frac_error})), which reaches a maximum of
$80\unit{\%}$ at $\eta_m=0$.

\section{General separation of the two components}

In the main text we have extracted a few parameters of the heat
capacity and analyzed them in terms of the superconducting mass
fraction. This has allowed us to determine the parameters for an
ideal sample at $\eta_m = 1$. We now want to extract the full
temperature dependence of the heat capacity of the two components
$C_1(T)$ and $C_2(T)$ from the measured data.

To do this, we rewrite Eq.~(\ref{eqn:hc_definition}):
\begin{equation}\label{eqn:hc_decomp}
  \Cp(T) = \eta_m C_1(T) + (1-\eta_m) C_2(T)
\end{equation}
and we expand the heat capacities into power series:
\begin{equation}\label{eqn:hc_series_expansion}
  \Cp(T) = \sum_k a_k T^k \qquad C_i(T) = \sum_k a_{i,k} T^k, \qquad i \in \{1,2\}.
\end{equation}
From inserting the expansion (\ref{eqn:hc_series_expansion}) into
Eq.~(\ref{eqn:hc_decomp}) follows that
\begin{equation}\label{eqn:hc_series_coeff}
  a_k = (a_{1,k} - a_{2,k})\eta_m + a_{2,k} \qquad \forall k,
\end{equation}
i.e.~all coefficients $a_k$ of the series expansion are linear in
$\eta_m$.

Therefore, a general recipe to separate the measured quantity into
its components is to expand it into a power series. The
coefficients of the power series depend linearly on the \smf\
$\eta_m$ and thus the intrinsic coefficients for component one
($a_{1,k}$) and two ($a_{2,k}$) can be extracted by a linear fit
when $a_k$ is plotted vs $\eta_m$. Using these coefficients, we
can easily reconstruct $C_1(T)$ and $C_2(T)$.

Of course, the measured Sommerfeld coefficient $\sfc$ together
with the intrinsic coefficients $\sfc_1$ and $\sfc_2$ which were
used in the main text are just a special case of the above
analysis with $a_{1} \equiv \sfc$, $a_{1,1} \equiv \sfc_1$, and
$a_{2,1} \equiv \sfc_2$.

\section{Alternative sample property measured}

As discussed in the main text, this analysis is necessary because
the thermodynamic quantity of interest is an \textit{intensive}
heat capacity. However, since the relative fraction of the two
phases is not known a priori, the mass (volume, a.o.s.) used to
convert the extensive heat capacity into intensive heat capacity
is the \textit{total} mass (volume, a.o.s.) of the sample
measured. This leads to an error in the heat capacity, which can
be eliminated using the condensation energy analysis given above.

The set of formulas for the condensation energy analysis depends
on the units the heat capacity is given in and the sample property
measured (mass, volume, or a.o.s.). Whereas mass is the most often
encountered case of a sample property measured, we also list the
formulas for other situations in Table \ref{tab:Other_forms}. In
the main text we have chosen to give the heat capacity per mass
because this simplifies the equations when the total mass is
measured. Considering our analysis, we propose that for practical
purposes heat capacity should always be specified per sample
property measured, i.e.~per mass in the most cases. A mathematical
conversion assuming a tabulated molar mass, for example, renders
data inscrutable. Such conversions should only be done for
parameters of an ideal sample.

\begin{table}[!]
\caption{\label{tab:Other_forms} Formula sets for the CEA
describing parameters as a function of the superconducting
fraction. Cases where the heat capacity is given per sample
property measured are given in the first three sets, from which
the first set is used in this article. $\eta_{n} = n_1/n$ denotes
the superconducting a.o.s.~fraction. The last set (measured mass,
heat capacity per a.o.s.) is the most frequently encountered
situation in publications and is therefore listed here as well. In
this case the formulas include the factor $\nu = M_1/M_2$ which is
the ratio of the molar masses of the two components. We use the
expansion $\Cp(T) = \sum_k a_k T^k$ and $C_i(T) = \sum_k a_{i,k}
T^k$, where $i \in \{1,2\}$, for the heat capacity, so e.g.~$a_{1}
\equiv \sfc$, $a_{1,1} \equiv \sfc_1$, and $a_{2,1} \equiv
\sfc_2$.}
\begin{ruledtabular}
\begin{tabular}{lcr}
CEA formulas & Sample property measured & Heat capacity per\\
\hline
$a_k = (a_{1,k} - a_{2,k})\eta_m + a_{2,k} \qquad \forall k$ \\
$\sfc_\mathrm{r} = -\sfc_2\eta_m + \sfc_2$ & mass & mass\\
$C_\mathrm{es} = \eta_m^{-1} \Delta \Cp + \sfc_1 T$\\
\hline
$a_k = (a_{1,k} - a_{2,k})\eta_V + a_{2,k} \qquad \forall k$ \\
$\sfc_\mathrm{r} = -\sfc_2\eta_V + \sfc_2$ & volume & volume\\
$C_\mathrm{es} = \eta_V^{-1} \Delta \Cp + \sfc_1 T$\\
\hline
$a_k = (a_{1,k} - a_{2,k})\eta_{n} + a_{2,k} \qquad \forall k$ \\
$\sfc_\mathrm{r} = -\sfc_2\eta_{n} + \sfc_2$ & a.o.s.\footnote{The
a.o.s.~of a two-component sample could be directly measured by a
titration process, with two titrating reagents matched to the two
components. However, this cumbersome process destroys the sample
and would therefore have to be performed \textit{after} the sample
has been measured. It further requires detailed knowledge of the
two components and is therefore unlikely to be used for our
purpose. We nevertheless include the formulas
for this situation for the sake of completeness.} & a.o.s.\\
$C_\mathrm{es} = \eta_{n}^{-1} \Delta \Cp + \sfc_1 T$\\
\hline \hline
$a_k = (a_{1,k} - \nu a_{2,k})\eta_m + \nu a_{2,k} \qquad \forall k$ \\
$\sfc_\mathrm{r} = -\nu\sfc_2\eta_m + \nu\sfc_2$ & mass & a.o.s.\\
$C_\mathrm{es} = \eta_m^{-1} \Delta \Cp + \sfc_1 T$\\
\end{tabular}
\end{ruledtabular}
\end{table}

\section{Inclusion of a $C_\mathrm{mag}$ term}\label{sec:Incl_Cmag}

In Section~\ref{sec:normal_props} of the main text we have
attributed the deviation of the $\Delta \Cp$-slope-line
$y$-intercept from zero to a finite magnetic heat capacity. Even
though the deviation of the $y$-intercept is fairly small and may
be attributed to measurement uncertainty, we nevertheless make an
attempt to explain it here. It is shown that a refined analysis
including a linear-in-$T$ $C_\mathrm{mag}$ term can account for
this behavior. In Section~\ref{sec:SC_props} we have mentioned
that a fine-tuned CEA including such a term is used for the
extraction of $C_\mathrm{es}(T)$ on a logarithmic scale, where the
behavior at low temperatures is much expanded. The details of this
refinement are given here as well. Further, it is shown that the
inclusion of such a term leaves the extracted parameters
practically unchanged.

To account for the magnetic heat capacity from various possible
sources like magnetic impurities, magnons, or fluctuations, we may
assume an approximate form $C_2^\mathrm{mag} = \sfc_2^\mathrm{mag}
T$ to first order. This is reasonable considering that e.g.~a
Schottky-type impurity has a maximum $\Cp/T$ contribution around
$2$ to $5\unit{K}$ at $12\unit{T}$, depending on the $g$-factor
and the interaction strength. In this temperature region the
extraction of both the Sommerfeld coefficient $\sfc$ and the slope
of $\Delta \Cp$ are performed. Thus, traces of impurities may
result in an additional "linear term" causing $\sfc$ and
$-\mathrm{d}(\Delta \Cp) / \mathrm{d}T$ to increase.

Traces of magnetic impurities have been identified in the starting
material and additional impurity phases like RbOsO$_4$ may be
produced by the synthesis procedure. We would like to emphasize
that such phases are present in the ppm range in vast contrast to
the $17$ to $66$ volume-percent OsO$_2$, the component relevant
for the CEA. Magnetic trace impurities are merely mentioned here
to motivate the inclusion of the additional small term
$\sfc_2^\mathrm{mag}$. As will be shown below, the extracted
parameters are essentially unchanged and thus the inclusion of
$\sfc_2^\mathrm{mag} T$ is a "second order" extension to the
analysis.

Including $C_2^\mathrm{mag} = \sfc_2^\mathrm{mag} T$ extends the
CEA as follows: Equations (\ref{eqn:gamma_b}) and
(\ref{eqn:DeltaC}) from the main text describing the $\sfc$- and
the $\Delta \Cp$-slope-line are modified slightly to
\begin{eqnarray}\label{eqn:CEA_incl_Cmag}
  \sfc = \left[\sfc_1 - (\sfc_2 + \sfc_2^\mathrm{mag})\right] \eta_m + (\sfc_2 + \sfc_2^\mathrm{mag}) \\
  -\mathrm{d}(\Delta \Cp) / \mathrm{d}T = (\sfc_1 - \sfc_2^\mathrm{mag})
  \eta_m + \sfc_2^\mathrm{mag},
\end{eqnarray}
while Eq.~(\ref{eqn:gamma_res_b}) describing the
$\sfc_\mathrm{r}$-line remains the same:
\begin{equation}\label{eqn:CEA_incl_Cmagb}
  \sfc_\mathrm{r} = - \sfc_2 \eta_m + \sfc_2.
\end{equation}
In other words, the $\sfc$- and the $\Delta \Cp$-slope-line are
changed in that their $y$-intercept is being moved up slightly by
$\sfc_2^\mathrm{mag}$.

\begin{figure}
  \includegraphics{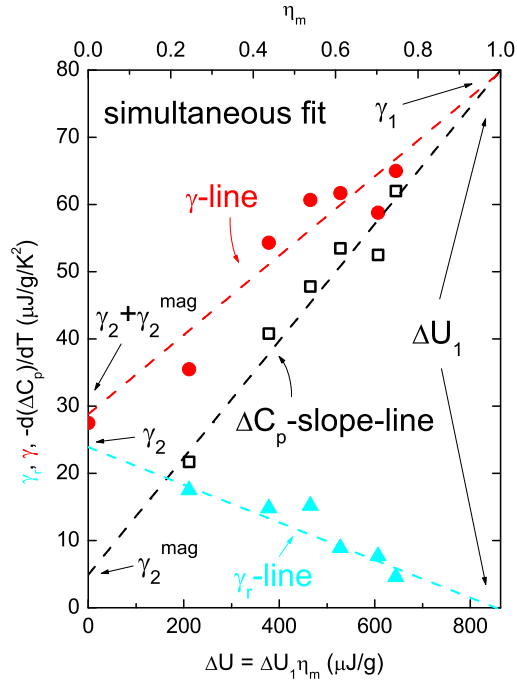}
  \caption{\label{fig:BruhwilerPlot_per_mass_simult} (Color online) Refined CEA
  including a $C_\mathrm{mag}$ term: The data are simultaneously
  fitted to the $\sfc$-, $\Delta \Cp$-slope-, and the
  $\sfc_\mathrm{r}$-line, see Eqs.~(\ref{eqn:CEA_incl_Cmag}) to (\ref{eqn:CEA_incl_Cmagb}).
  The result is essentially the same
  as when the lines are fitted separately (cf.~Fig.~\ref{fig:BruhwilerPlot}).}
\end{figure}

A simultaneous fit of all three lines using $\sfc_1$, $\sfc_2$,
$\sfc_2^\mathrm{mag}$, and $\ce_1$ as fit parameters, results in
parameters identical within experimental error to the results from
the analysis without a $C_2^\mathrm{mag}$ term: $\sfc_1$ increases
by $0.6\unit{\%}$ to $79.6\unit{\mu J/g/K^2}$ and $\ce_1$
increases by $0.3\unit{\%}$ to $863\unit{\mu J/g}$.
$\sfc_2^\mathrm{mag}$ turns out to be $4.9\unit{\mu J/g/K^2}$,
i.e.~about $6\unit{\%}$ of the normal state Sommerfeld coefficient
$\sfc_1$. The refined fit is shown in
Fig.~\ref{fig:BruhwilerPlot_per_mass_simult}.

\begin{figure}
  \includegraphics{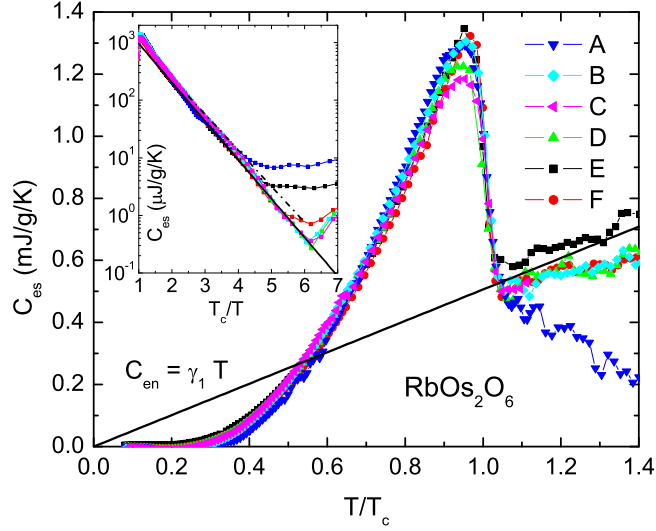}
  \caption{\label{fig:Ces_vs_TdivTc} (Color online) Superconducting electronic specific
   heat $C_\mathrm{es}(T) = \eta_m^{-1} \Delta \Cp +
  \sfc_1 T$ for all measured samples showing a universal behavior
  when scaled appropriately. The inset shows the
superconducting electronic specific heat of \fu\ showing close
  quantitative agreement with conventional superconducting
  behavior. The solid line is a best fit to the data and indicates
  the expected behavior from an isotropic gap:
$C_\mathrm{es} = 9 \, \sfc_1 \, T_\mathrm{c} \exp(-1.55 \,
T_\mathrm{c}/T)$. For comparison the dash-dotted line shows the
expected behavior from weak coupling BCS. The deviations at very
low temperatures are due to scattering from various minute
impurities.}
\end{figure}

The superconducting electronic heat capacity $C_\mathrm{es}(T)$ on
a global scale is well represented by the extracted $\sfc_1$ using
Eq.~\ref{eqn:Ces} (cf.~Fig.~\ref{fig:Ces_vs_TdivTc}). Plotting
$C_\mathrm{es}(T)$ on a logarithmic scale, however, emphasizes the
lowest temperatures and it is thus important to include the
effects of the $\sfc_2^\mathrm{mag}$ term:

\begin{equation}\label{eqn:Ces_incl_Cmag}
  C_\mathrm{es}(T) = \eta_m^{-1} \Delta \Cp + \left[ \sfc_1 + (\eta_m^{-1}-1)\sfc_2^\mathrm{mag} \right]
  T.
\end{equation}
Based on this, we have chosen effective Sommerfeld coefficients
for the calculation of $C_\mathrm{es}(T)$ by optimizing the
congruence among the curves (inset of
Fig.~\ref{fig:Ces_vs_TdivTc}). The so-obtained coefficients follow
the general trend $\sfc_1 + (\eta_m^{-1}-1)\sfc_2^\mathrm{mag}$,
showing that the inclusion of $\sfc_2^\mathrm{mag} T$ is
reasonable. It also becomes clear that magnetic impurities have a
significant effect on the extraction of $C_\mathrm{es}(T)$ at
small \svf s and low temperatures.

\newpage


\begin{thebibliography}{17}
\expandafter\ifx\csname
natexlab\endcsname\relax\def\natexlab#1{#1}\fi
\expandafter\ifx\csname bibnamefont\endcsname\relax
  \def\bibnamefont#1{#1}\fi
\expandafter\ifx\csname bibfnamefont\endcsname\relax
  \def\bibfnamefont#1{#1}\fi
\expandafter\ifx\csname citenamefont\endcsname\relax
  \def\citenamefont#1{#1}\fi
\expandafter\ifx\csname url\endcsname\relax
  \def\url#1{\texttt{#1}}\fi
\expandafter\ifx\csname
urlprefix\endcsname\relax\def\urlprefix{URL }\fi
\providecommand{\bibinfo}[2]{#2}
\providecommand{\eprint}[2][]{\url{#2}}

\bibitem[{\citenamefont{Hanawa et~al.}(2001)\citenamefont{Hanawa, Muraoka,
  Tayama, Sakakibara, Yamaura, and Hiroi}}]{HaMuTaSaYaHi2001}
\bibinfo{author}{\bibfnamefont{M.}~\bibnamefont{Hanawa}},
  \bibinfo{author}{\bibfnamefont{Y.}~\bibnamefont{Muraoka}},
  \bibinfo{author}{\bibfnamefont{T.}~\bibnamefont{Tayama}},
  \bibinfo{author}{\bibfnamefont{T.}~\bibnamefont{Sakakibara}},
  \bibinfo{author}{\bibfnamefont{J.}~\bibnamefont{Yamaura}}, \bibnamefont{and}
  \bibinfo{author}{\bibfnamefont{Z.}~\bibnamefont{Hiroi}},
  \bibinfo{journal}{\prl} \textbf{\bibinfo{volume}{87}},
  \bibinfo{pages}{187001} (\bibinfo{year}{2001}).

\bibitem[{\citenamefont{Sakai et~al.}(2001)\citenamefont{Sakai, Yoshimura,
  Ohno, Kato, Kambe, Walstedt, Matsuda, Haga, and
  Onuki}}]{SaYoOhKaKaWaMaHaOn2001}
\bibinfo{author}{\bibfnamefont{H.}~\bibnamefont{Sakai}},
  \bibinfo{author}{\bibfnamefont{K.}~\bibnamefont{Yoshimura}},
  \bibinfo{author}{\bibfnamefont{H.}~\bibnamefont{Ohno}},
  \bibinfo{author}{\bibfnamefont{H.}~\bibnamefont{Kato}},
  \bibinfo{author}{\bibfnamefont{S.}~\bibnamefont{Kambe}},
  \bibinfo{author}{\bibfnamefont{R.~E.} \bibnamefont{Walstedt}},
  \bibinfo{author}{\bibfnamefont{T.~D.} \bibnamefont{Matsuda}},
  \bibinfo{author}{\bibfnamefont{Y.}~\bibnamefont{Haga}}, \bibnamefont{and}
  \bibinfo{author}{\bibfnamefont{Y.}~\bibnamefont{Onuki}}, \bibinfo{journal}{J.
  Phys.: Condens. Matter} \textbf{\bibinfo{volume}{13}}, \bibinfo{pages}{L785}
  (\bibinfo{year}{2001}).

\bibitem[{\citenamefont{Jin et~al.}(2001)\citenamefont{Jin, He, McCall,
  Alexander, Drymiotis, and Mandrus}}]{JiHeMcAlDrMa2001}
\bibinfo{author}{\bibfnamefont{R.}~\bibnamefont{Jin}},
  \bibinfo{author}{\bibfnamefont{J.}~\bibnamefont{He}},
  \bibinfo{author}{\bibfnamefont{S.}~\bibnamefont{McCall}},
  \bibinfo{author}{\bibfnamefont{C.~S.} \bibnamefont{Alexander}},
  \bibinfo{author}{\bibfnamefont{F.}~\bibnamefont{Drymiotis}},
  \bibnamefont{and} \bibinfo{author}{\bibfnamefont{D.}~\bibnamefont{Mandrus}},
  \bibinfo{journal}{\prb} \textbf{\bibinfo{volume}{64}},
  \bibinfo{pages}{180503} (\bibinfo{year}{2001}).

\bibitem[{\citenamefont{Yonezawa
  et~al.}(2004{\natexlab{a}})\citenamefont{Yonezawa, Muraoka, and
  Hiroi}}]{YoMuHi2004}
\bibinfo{author}{\bibfnamefont{S.}~\bibnamefont{Yonezawa}},
  \bibinfo{author}{\bibfnamefont{Y.}~\bibnamefont{Muraoka}}, \bibnamefont{and}
  \bibinfo{author}{\bibfnamefont{Z.}~\bibnamefont{Hiroi}}, \bibinfo{journal}{J.
  Phys. Soc. Jpn.} \textbf{\bibinfo{volume}{73}}, \bibinfo{pages}{1655}
  (\bibinfo{year}{2004}{\natexlab{a}}).

\bibitem[{\citenamefont{Yonezawa
  et~al.}(2004{\natexlab{b}})\citenamefont{Yonezawa, Muraoka, Matsushita, and
  Hiroi}}]{YoMuMaHi2004a}
\bibinfo{author}{\bibfnamefont{S.}~\bibnamefont{Yonezawa}},
  \bibinfo{author}{\bibfnamefont{Y.}~\bibnamefont{Muraoka}},
  \bibinfo{author}{\bibfnamefont{Y.}~\bibnamefont{Matsushita}},
  \bibnamefont{and} \bibinfo{author}{\bibfnamefont{Z.}~\bibnamefont{Hiroi}},
  \bibinfo{journal}{J. Phys. Soc. Jpn.} \textbf{\bibinfo{volume}{73}},
  \bibinfo{pages}{819} (\bibinfo{year}{2004}{\natexlab{b}}).

\bibitem[{\citenamefont{Yonezawa
  et~al.}(2004{\natexlab{c}})\citenamefont{Yonezawa, Muraoka, Matsushita, and
  Hiroi}}]{YoMuMaHi2004}
\bibinfo{author}{\bibfnamefont{S.}~\bibnamefont{Yonezawa}},
  \bibinfo{author}{\bibfnamefont{Y.}~\bibnamefont{Muraoka}},
  \bibinfo{author}{\bibfnamefont{Y.}~\bibnamefont{Matsushita}},
  \bibnamefont{and} \bibinfo{author}{\bibfnamefont{Z.}~\bibnamefont{Hiroi}},
  \bibinfo{journal}{J. Phys.: Condens. Matter} \textbf{\bibinfo{volume}{16}},
  \bibinfo{pages}{L9} (\bibinfo{year}{2004}{\natexlab{c}}).

\bibitem[{\citenamefont{Br{\"u}hwiler et~al.}(2004)\citenamefont{Br{\"u}hwiler,
  Kazakov, Zhigadlo, Karpinski, and Batlogg}}]{BrKaZhKaBa2004}
\bibinfo{author}{\bibfnamefont{M.}~\bibnamefont{Br{\"u}hwiler}},
  \bibinfo{author}{\bibfnamefont{S.~M.} \bibnamefont{Kazakov}},
  \bibinfo{author}{\bibfnamefont{N.~D.} \bibnamefont{Zhigadlo}},
  \bibinfo{author}{\bibfnamefont{J.}~\bibnamefont{Karpinski}},
  \bibnamefont{and} \bibinfo{author}{\bibfnamefont{B.}~\bibnamefont{Batlogg}},
  \bibinfo{journal}{\prb} \textbf{\bibinfo{volume}{70}},
  \bibinfo{pages}{020503(R)} (\bibinfo{year}{2004}).

\bibitem[{\citenamefont{Khasanov et~al.}(2004)\citenamefont{Khasanov,
  Eshchenko, Karpinski, Kazakov, Zhigadlo, Br{\"u}tsch, Gavillet, {Di Castro},
  Shengelaya, {La Mattina} et~al.}}]{KhEsKaKaZhBrGaDiShMaMaBaKe2004}
\bibinfo{author}{\bibfnamefont{R.}~\bibnamefont{Khasanov}},
  \bibinfo{author}{\bibfnamefont{D.~G.} \bibnamefont{Eshchenko}},
  \bibinfo{author}{\bibfnamefont{J.}~\bibnamefont{Karpinski}},
  \bibinfo{author}{\bibfnamefont{S.~M.} \bibnamefont{Kazakov}},
  \bibinfo{author}{\bibfnamefont{N.~D.} \bibnamefont{Zhigadlo}},
  \bibinfo{author}{\bibfnamefont{R.}~\bibnamefont{Br{\"u}tsch}},
  \bibinfo{author}{\bibfnamefont{D.}~\bibnamefont{Gavillet}},
  \bibinfo{author}{\bibfnamefont{D.}~\bibnamefont{{Di Castro}}},
  \bibinfo{author}{\bibfnamefont{A.}~\bibnamefont{Shengelaya}},
  \bibinfo{author}{\bibfnamefont{F.}~\bibnamefont{{La Mattina}}},
  \bibnamefont{et~al.}, \bibinfo{journal}{\prl} \textbf{\bibinfo{volume}{93}},
  \bibinfo{pages}{157004} (\bibinfo{year}{2004}).

\bibitem[{\citenamefont{Magishi et~al.}()\citenamefont{Magishi, Gavilano,
  Pedrini, Hinderer, Weller, Ott, Kazakov, and
  Karpinski}}]{MaGaPeHiWeOtKaKa2004}
\bibinfo{author}{\bibfnamefont{K.}~\bibnamefont{Magishi}},
  \bibinfo{author}{\bibfnamefont{J.~L.} \bibnamefont{Gavilano}},
  \bibinfo{author}{\bibfnamefont{B.}~\bibnamefont{Pedrini}},
  \bibinfo{author}{\bibfnamefont{J.}~\bibnamefont{Hinderer}},
  \bibinfo{author}{\bibfnamefont{M.}~\bibnamefont{Weller}},
  \bibinfo{author}{\bibfnamefont{H.~R.} \bibnamefont{Ott}},
  \bibinfo{author}{\bibfnamefont{S.~M.} \bibnamefont{Kazakov}},
  \bibnamefont{and}
  \bibinfo{author}{\bibfnamefont{J.}~\bibnamefont{Karpinski}},
  \eprint{cond-mat/0409169}.

\bibitem[{\citenamefont{Kazakov et~al.}(2004)\citenamefont{Kazakov, Zhigadlo,
  Br{\"u}hwiler, Batlogg, and Karpinski}}]{KaZhBrBaKa2004}
\bibinfo{author}{\bibfnamefont{S.~M.} \bibnamefont{Kazakov}},
  \bibinfo{author}{\bibfnamefont{N.~D.} \bibnamefont{Zhigadlo}},
  \bibinfo{author}{\bibfnamefont{M.}~\bibnamefont{Br{\"u}hwiler}},
  \bibinfo{author}{\bibfnamefont{B.}~\bibnamefont{Batlogg}}, \bibnamefont{and}
  \bibinfo{author}{\bibfnamefont{J.}~\bibnamefont{Karpinski}},
  \bibinfo{journal}{Supercond. Sci. Technol.} \textbf{\bibinfo{volume}{17}},
  \bibinfo{pages}{1169} (\bibinfo{year}{2004}).

\bibitem[{\citenamefont{Marsiglio et~al.}(1987)\citenamefont{Marsiglio,
  Coombes, and Carbotte}}]{MaCoCa1987}
\bibinfo{author}{\bibfnamefont{F.}~\bibnamefont{Marsiglio}},
  \bibinfo{author}{\bibfnamefont{J.~M.} \bibnamefont{Coombes}},
  \bibnamefont{and} \bibinfo{author}{\bibfnamefont{J.~P.}
  \bibnamefont{Carbotte}}, \bibinfo{journal}{\prb}
  \textbf{\bibinfo{volume}{35}}, \bibinfo{pages}{3219} (\bibinfo{year}{1987}).

\bibitem[{\citenamefont{Kunes et~al.}(2004)\citenamefont{Kunes, Jeong, and
  Pickett}}]{KuJePi2004}
\bibinfo{author}{\bibfnamefont{J.}~\bibnamefont{Kunes}},
  \bibinfo{author}{\bibfnamefont{T.}~\bibnamefont{Jeong}}, \bibnamefont{and}
  \bibinfo{author}{\bibfnamefont{W.~E.} \bibnamefont{Pickett}},
  \bibinfo{journal}{Physical Review B (Condensed Matter and Materials Physics)}
  \textbf{\bibinfo{volume}{70}}, \bibinfo{eid}{174510}
  (pages~\bibinfo{numpages}{6}) (\bibinfo{year}{2004}),
  \urlprefix\url{http://link.aps.org/abstract/PRB/v70/e174510}.

\bibitem[{\citenamefont{Saniz et~al.}(2004)\citenamefont{Saniz, Medvedeva, Ye,
  Shishidou, and Freeman}}]{SaMeYeShFr2004}
\bibinfo{author}{\bibfnamefont{R.}~\bibnamefont{Saniz}},
  \bibinfo{author}{\bibfnamefont{J.~E.} \bibnamefont{Medvedeva}},
  \bibinfo{author}{\bibfnamefont{L.-H.} \bibnamefont{Ye}},
  \bibinfo{author}{\bibfnamefont{T.}~\bibnamefont{Shishidou}},
  \bibnamefont{and} \bibinfo{author}{\bibfnamefont{A.~J.}
  \bibnamefont{Freeman}}, \bibinfo{journal}{\prb}
  \textbf{\bibinfo{volume}{70}}, \bibinfo{pages}{100505(R)}
  (\bibinfo{year}{2004}).

\bibitem[{\citenamefont{Schneider et~al.}()\citenamefont{Schneider, Khasanov,
  and Keller}}]{SchKhKe2004}
\bibinfo{author}{\bibfnamefont{T.}~\bibnamefont{Schneider}},
  \bibinfo{author}{\bibfnamefont{R.}~\bibnamefont{Khasanov}}, \bibnamefont{and}
  \bibinfo{author}{\bibfnamefont{H.}~\bibnamefont{Keller}},
  \eprint{cond-mat/0409398}.

\bibitem[{\citenamefont{Marsiglio and Carbotte}(1986)}]{MaCa1986}
\bibinfo{author}{\bibfnamefont{F.}~\bibnamefont{Marsiglio}} \bibnamefont{and}
  \bibinfo{author}{\bibfnamefont{J.~P.} \bibnamefont{Carbotte}},
  \bibinfo{journal}{\prb} \textbf{\bibinfo{volume}{33}}, \bibinfo{pages}{6141}
  (\bibinfo{year}{1986}).

\bibitem[{\citenamefont{Bardeen and Schrieffer}(1961)}]{BaSch1961}
\bibinfo{author}{\bibfnamefont{J.}~\bibnamefont{Bardeen}} \bibnamefont{and}
  \bibinfo{author}{\bibfnamefont{J.~R.} \bibnamefont{Schrieffer}},
  \emph{\bibinfo{title}{Recent Developments in Superconductivity}}, vol.
  \bibinfo{volume}{III} of \emph{\bibinfo{series}{Progress in low temperature
  physics}} (\bibinfo{publisher}{north-holland}, \bibinfo{year}{1961}).

\bibitem[{\citenamefont{Leupold and Boorse}(1964)}]{LeBo1964}
\bibinfo{author}{\bibfnamefont{H.~A.} \bibnamefont{Leupold}} \bibnamefont{and}
  \bibinfo{author}{\bibfnamefont{H.~A.} \bibnamefont{Boorse}},
  \bibinfo{journal}{Phys. Rev.} \textbf{\bibinfo{volume}{134}},
  \bibinfo{pages}{A1322} (\bibinfo{year}{1964}).

\end{thebibliography}
\end{document}